\documentclass[manuscript]{acmart} 
\AtBeginDocument{%
  \providecommand\BibTeX{{%
    \normalfont B\kern-0.5em{\scshape i\kern-0.25em b}\kern-0.8em\TeX}}}

\setcopyright{acmcopyright}

\usepackage{amsmath}
\usepackage{textcomp}
\usepackage{booktabs}
\usepackage{graphicx}
\usepackage{subcaption}
\usepackage{multirow}
\usepackage[inline]{enumitem}
\usepackage[justification=justified]{caption}
\makeatletter  
\newif\if@restonecol  
\makeatother

\makeatletter  
\newif\if@restonecol  
\makeatother

\usepackage[linesnumbered,ruled,vlined]{algorithm2e}
\usepackage{algpseudocode} 
\usepackage{listings}
\definecolor{delim}{RGB}{20,105,176}
\definecolor{numb}{RGB}{106, 109, 32}
\definecolor{string}{rgb}{0.64,0.08,0.08}
\usepackage{pifont}

\newcommand{\cmark}{\ding{51}}%
\newcommand{\xmark}{\ding{55}}%

\lstdefinelanguage{json}{
    numbers=left,
    numberstyle=\small,
    frame=single,
    rulecolor=\color{black},
    showspaces=false,
    showtabs=false,
    breaklines=true,
    postbreak=\raisebox{0ex}[0ex][0ex]{\ensuremath{\color{gray}\hookrightarrow\space}},
    breakatwhitespace=true,
    basicstyle=\ttfamily\footnotesize\linespread{0.8},
    upquote=true,
    morestring=[b]",
    stringstyle=\color{string},
    literate=
     *{0}{{{\color{numb}0}}}{1}
      {1}{{{\color{numb}1}}}{1}
      {2}{{{\color{numb}2}}}{1}
      {3}{{{\color{numb}3}}}{1}
      {4}{{{\color{numb}4}}}{1}
      {5}{{{\color{numb}5}}}{1}
      {6}{{{\color{numb}6}}}{1}
      {7}{{{\color{numb}7}}}{1}
      {8}{{{\color{numb}8}}}{1}
      {9}{{{\color{numb}9}}}{1}
      {\{}{{{\color{delim}{\{}}}}{1}
      {\}}{{{\color{delim}{\}}}}}{1}
      {[}{{{\color{delim}{[}}}}{1}
      {]}{{{\color{delim}{]}}}}{1},
}

\usepackage{todonotes}
\def\BibTeX{{\rm B\kern-.05em{\sc i\kern-.025em b}\kern-.08em
    T\kern-.1667em\lower.7ex\hbox{E}\kern-.125emX}}
\begin{document}

\newcommand{\revisedmajor}[2]{\color{black}#1 #2\color{black}\xspace}
\newcommand{\setredmajor}{\color{black}}






\title{A Decentralized and Self-Adaptive Approach for Monitoring Volatile Edge Environments}
%



\author{Shashikant Ilager}
\affiliation{%
  \institution{Vienna University of Technology}
  \city{Vienna}
  \country{Austria}}
\email{shashikant.ilager@tuwien.ac.at}
\orcid{0000-0003-1178-6582}

\author{Jakob Fahringer}
\affiliation{%
  \institution{Vienna University of Technology}
  \city{Vienna}
  \country{Austria}}
\email{jakob.fahringer@tuwien.ac.at}

\author{Alessandro Tundo}
\affiliation{%
 \institution{University of Milano-Bicocca}
 \city{Milan}
 \country{Italy}}
\email{alessandro.tundo@unimib.it}
\orcid{0000-0001-8840-8948}

\author{Ivona Brandić}
\affiliation{%
  \institution{Vienna University of Technology}
  \city{Vienna}
  \country{Austria}}
\email{ivona.brandic@tuwien.ac.at}
\orcid{0000-0001-7424-0208}

\newcommand{\repo}{\url{https://github.com/hpc-tuwien/DEMon/}\xspace}

\begin{abstract}
Edge computing provides resources for IoT workloads at the network edge. Monitoring systems are vital   for efficiently managing resources and application workloads by collecting, storing, and providing relevant information about the state of the resources. However, traditional monitoring systems have a centralized architecture for both data plane and control plane, which increases latency, creates a failure bottleneck, and faces challenges in providing quick and trustworthy data in volatile edge environments, especially where infrastructures are often built upon failure-prone, unsophisticated computing and network resources. Thus, we propose DEMon, a decentralized, self-adaptive monitoring system for edge. DEMon leverages the stochastic gossip communication protocol at its core. It develops efficient protocols for information dissemination, communication, and retrieval, avoiding a single point of failure and ensuring fast and trustworthy data access. Its decentralized control enables self-adaptive management of monitoring parameters, addressing the trade-offs between the quality of service of monitoring and resource consumption. We implement the proposed system as a lightweight and portable container-based system and evaluate it through experiments. We also present a use case demonstrating its feasibility. The results show that DEMon efficiently disseminates and retrieves the monitoring information, addressing the challenges of edge monitoring.
 
\end{abstract}



\maketitle


\section{Introduction}
\label{sec:introduction}

Edge computing offers computing resources for the latency-sensitive Internet of Things (IoT) workloads, enabling the data processing at the network edge~\cite{shi2016edge,satyanarayanan2017emergence}.   Unlike Cloud Computing, which provides reliable and robust computing resources from centralized data centers, edge computing offers services from a highly distributed environment with heterogeneous and resource-constrained compute and network resources. Despite its shortcomings, the massive demand for latency-sensitive and time-critical applications such as smart cities and autonomous vehicles necessitates the widespread deployment of edge platforms~\cite{buyya2018manifesto, satyanarayanan2019computing}. In this regard, deploying application services reliably on edge needs efficient infrastructure monitoring systems, allowing applications and service providers to make crucial decisions based on the monitoring data. 

\par Monitoring services allow observation of the overall status of the infrastructure and play a crucial role in resource management tasks such as resource provisioning, scheduling, load balancing, and failure detection. Traditionally, cloud services are offered through multiple data centres directly managed by single service providers. Cloud data center resources are monitored through sophisticated Data Center Infrastructure Management tools centrally deployed on robust and reliable servers. Many of the cloud service providers build their in-house platforms (e.g., Google's Borgmon), while private clouds adopt open-source monitoring tools such as Zabbix~\cite{olups2010zabbix} and Prometheus~\cite{turnbull2018monitoring}. In all of such monitoring systems, both the \emph{data plane} and \emph{control plane}, possess centralized architecture.
The \emph{control plane} fine-tunes the configuration parameters such as data collection and forwarding intervals, parameters selection, and other application and infrastructure-specific configurations. A centralized system pushes the configuration rules to all the infrastructure nodes. On the other hand, the \emph{data plane} is responsible for storing the monitoring data, i.e., the state of other servers. Such monitoring data is collected on predefined time intervals and usually stored in time-series databases hosted on centralized servers. However, such monitoring systems require dedicated and reliable computational and storage resources, which are infeasible at the edge due to their unique requirements and challenges. 

\par \emph{First,} edge Computing infrastructure is highly heterogeneous, consisting of IoT devices with onboard CPUs, embedded systems, domain-specific accelerators~\cite{park2020mobile}, and inexpensive off-the-shelf commodity servers and micro-data centres. Such hyper-heterogeneity with resource-constrained, failure-prone devices introduces a massive complexity to the design and implementation of monitoring systems, where a single configuration would not work for all types of nodes. \emph{Secondly}, unlike a Cloud with a high-speed and reliable network, edge infrastructures are built upon limited bandwidth and unreliable networks, including wireless and cellular networks. Thus, communication failures should be considered a norm rather than an exception. \emph{Thirdly}, edge infrastructures are volatile, where resources are pooled by multiple service providers across multiple network domains~\cite{buyya2018manifesto}, and machines are dynamically provisioned or de-provisioned (join and leave the resource pool) based on network connectivity and power budget, among other parameters. This further challenges retrieving trustable monitoring data since resources belong to multiple parties~\cite{trust1_wang2020edge,trust2_FORTI2020775}.
Therefore, we require a monitoring system that is completely decentralized for both \emph{control} and \emph{data plane}, where each participating node can control and fine-tune its parameters based on its capability and conditions. Despite some recent works have explored solutions for edge computing monitoring~\cite{ grogman_2017, Taherizadeh_2017, brandon2018fmone} and multi-tier Fog Computing~\cite{adaptievemon, fogmon}, they consider either some form of a centralized controller or remote storage mechanisms.

\par In this paper, we propose an \emph{decentralized and self-adaptive monitoring system (DEMon)} for a highly-volatile edge environments. We envision the proposed monitoring system as distributed information management with efficient information spreading, storage, and data retrieval mechanisms. It provides a decentralized control and data storage method for edge monitoring system. We use a \emph{stochastic group communication protocol} for information dissemination in a volatile edge environment~\cite{kermarrec2007gossiping, birman2007promise, van2017epistemic}.

DEMon leverages a \emph{gossip-based information dissemination algorithm} capable of controlling according to the requirements of edge environments. \revisedmajor{}{The inherent stochastic characteristics of gossip-based protocol help create an adaptive and robust communication overlay network. This enables an effective decimation of information across the network in a decentralized manner without introducing massive concentrated network traffic on a specific network path and achieves a uniform network load distribution across the network.}

In addition, we propose the \emph{Leaderless Quorum Consensus (LQC)} protocol for information retrieval, which can quickly aggregate the information of a specific node, ensuring fast and trustworthy retrieval of the data. The DEMon's architecture is \emph{decentralized}, that is, \emph{it does not depend on a centralized controller and does not use centralized servers for information storage and retrieval}; instead, it uniformly distributes the data across the network autonomously. Furthermore, it is \emph{self-adaptive}, that is, no external configurations or measures are enforced during resource or network failures and infrastructure changes.  Moreover, application services can timely access monitoring data without increasing latency like centralized monitoring systems.

This work extends our preliminary short paper~\cite{ilagerdemon_UCC22} by: (i) providing a rigorous presentation of the monitoring system architecture; (ii) proposing a new state repository and network management techniques for efficient node insertion and deletion in the monitoring system; (iii) reporting results from an extensive empirical evaluation of the effectiveness of the approach and comparing it with a baseline; and (iv) showcasing the feasibility of the approach by simulating a real-world use case on a in-lab edge testbed.

In a nutshell, this paper provides the following contributions.

\textbf{A decentralized and self-adaptive monitoring system for highly-volatile edge environments.} We propose DEMon, a self-adaptive trustworthy monitoring system for highly-volatile edge environments that provides efficient and decentralized information spreading.

\textbf{A stochastic group communication protocol.} We present a gossip-based information dissemination algorithm and we study its hyper-parameters to understand their impact on edge environments monitoring.

\textbf{An efficient and trustworthy data retrieval method.} We propose the Leaderless Quorum Consensus (LQC) protocol for information retrieval, which can quickly aggregate trustable information of a specific node.

\textbf{A publicly available lightweight prototype.} We provide a publicly accessible prototype implementation of the DEMon monitoring system and algorithms.

\textbf{Empirical evidence of the effectiveness of the approach.} We answer four research questions by performing experiments in an emulated container-based environment.  Results show that DEMon can efficiently spread and retrieve monitoring information across edge networks, in a scalable manner.

\textbf{An use case implementation of a mobile computing edge scenario.} We implement a mobile computing use case employing 12 RaspberryPi devices representing different edge nodes deployed in an urban area. Each node hosts both an edge AI-based application (i.e., object detection application) and the DEMon components. We execute the use case and collect feasibility evidence of the proposed monitoring system by using real data belonging to the edge User Allocation~\cite{lai2018optimal_mel_dataset} dataset which has the recorded location entries of edge servers and mobile users.

The rest of the paper is organized as follows: Section~\ref{sec:motivation-and-background} introduces the use case scenario motivating our solution approach and provides the required background details. Section~\ref{sec:approach} presents the architecture of DEMon, its components, and the main algorithms. Section~\ref{sec:implementation} describes the implementation of the DEMon prototype. Section~\ref{sec:evaluation} presents the empirical evaluation and shows experimental results. Section~\ref{sec:related-work} discusses the relevant related works. Finally, Section~\ref{sec:conclusions} presents concluding remarks and future directions.

\section{Motivation and Background}
\label{sec:motivation-and-background}

\subsection{A Use Case for a Decentralized Edge Monitoring System}
\label{sec:motivation}

\begin{figure}[!ht]
\centering
\includegraphics[width=0.9\textwidth]{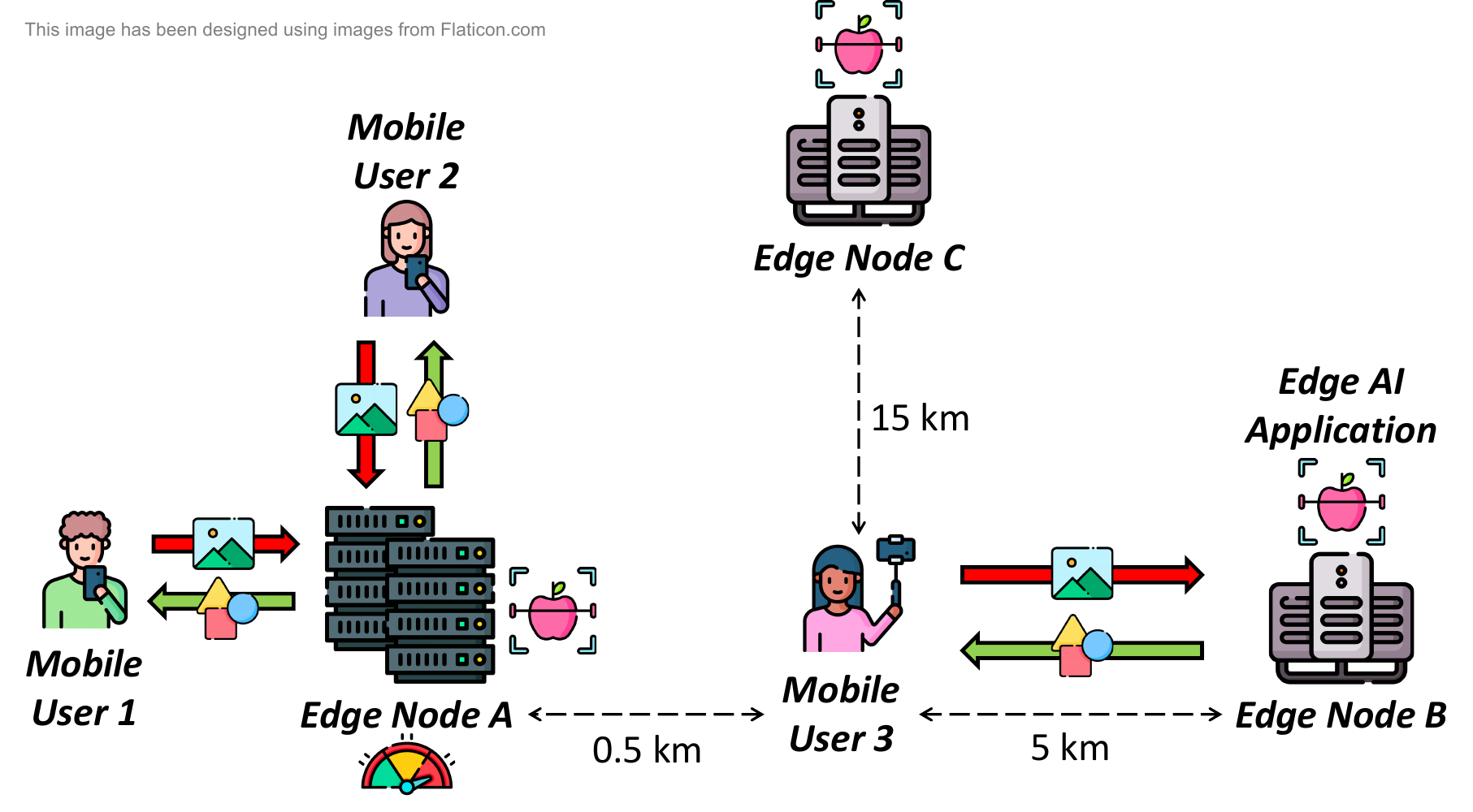}
\caption{An edge use case where mobile users offload a computational task to the nearest and less loaded edge node hosting an AI-based application.}
\label{fig:scenario} 
\end{figure}

Let us introduce a mobile computing use case to illustrate the requirements of a decentralized and self-adaptive monitoring system for the Edge computing. Figure~\ref{fig:scenario} depicts mobile users necessitating to offload their computational tasks (e.g., image object detection) to the nearby instances of an AI-based application. We consider that different resource provider dynamically pool their edge resources, and both the resource discovery protocol and the application instances are present on the edge nodes. Such application use cases using the edge nodes and AI-based applications can be found in many real-world applications such as environmental surveillance  \footnote{https://sagecontinuum.org/docs/about/overview} for fire hazard detection and  wildlife tracking,   air and water quality monitoring \footnote{https://www.chistera.eu/projects/swain}, and smart traffic management systems \footnote{https://intrasafed.ec.tuwien.ac.at},  among others. Therefore, a decentralized monitoring system is necessary for efficient information dissemination, storage, and querying in resource-constrained and volatile edge environments. 

Offloading task execution to edge nodes should be done with strict latency requirements (i.e., sub-milliseconds ) to timely satisfy user needs and respect QoS requirements 
~\cite{swarmdroneoffload, edge_object_detection_offloading}. For instance, to offload image processing to a nearby edge application instance, a mobile user has to find a suitable node that can accommodate their request by querying and aggregating information such as nodes availability, network and processing delay, and cost.  In such cases, a centralized monitoring system is not feasible due to increased latency and failure-prone edge environments ~\cite{fogmon, adaptievemon}.

In contrast, a mobile user can query the nearby edge nodes based on its current location, and obtain monitoring data that can help in finding a suitable node for their task. It must be considered this is possible only if the monitoring system is distributed, and the overall system's total or partial monitoring data is present in the nearby edge nodes. Moreover, providing trustable data is also required if we distribute the monitored data in an environment where multiple parties pool their resources to create a shared execution environment.

\subsection{Gossip Protocol}
\label{sec:gossip}

The epistemic algorithms are prominently used in many networked systems~\cite{van2017epistemic} such as for distributed database replication, data aggregation, and data clustering ~\cite{demers1987epidemic, van2003astrolabe, data_clusterazimi2018decentralized}. Epistemic algorithms are inherently stochastic and provide a robust framework for building communication protocols and information systems~\cite{van2017epistemic, birman2007promise}. The \emph{Gossip protocol} is a popular epistemic-based algorithm that provides efficient means for group communication without broadcasting. It is highly scalable and resilient, and it avoids a single point of failure~\cite{van1998gossip, kermarrec2007gossiping}.

The Gossip protocol itself is straightforward. Periodically, each node randomly selects a few other nodes, exchanges the state information, and waits to receive data from other nodes. The protocol provides a framework to dynamically control communication efficiency and resource usage, while it manages trade-offs between them. For instance, the rate of message exchange (\texttt{gossip\_rate}) and the number of random nodes chosen (\texttt{gossip\_count}) are configurable parameters, which affect the overall performance of the protocol. Once a node receives state information, it updates its state based on whether or not the data is already present. If received data is already present or older than the current data (based on timestamp), it drops the message and waits for new messages. If all the nodes know about every other node, it is confirmed that the system is converged. However, in our case, since the monitored data also changes continuously at each node, the gossiping continues indefinitely based on monitoring time intervals. 

\par It has been shown that the Gossip protocol works well in designing the theoretical distributed systems~\cite{birman2007promise}. However, its application in large-scale real-world systems is less explored~\cite{ditmarsch2016parameters}. In addition, if the hyper-parameters of the protocol, such as \texttt{gossip\_rate} and \texttt{gossip\_count} are misconfigured, it might exponentially increase the network and storage load, and it may even perform worse than broadcast-based communication~\cite{apt2017gossipcomputational}. Therefore, it becomes imperative to carefully study the feasibility of this stochastic method and analyze its effect on resource consumption, especially in resource-constrained edge environments. Consequently, in this paper, we study the effect of various hyper-parameters on metrics such as network and storage usage, speed of information propagation, query latency, and information quality.   


\section{DEMon: Decentralised Edge Monitoring and Control}
\label{sec:approach}

The DEMon monitoring system is designed following four main principles:

\begin{enumerate}

\item \textbf{Self-adaptation and configuration.} A system is self-adaptive and configurable if it can control and operate autonomously under dynamic conditions~\cite{self_adapt_christian_pmc15}. In that regard, DEMon does not depend on other nodes for control instructions and has no bootstrap configurations when new nodes join the system. Each node only sends and receives monitoring information through an independent DEMon agent that runs autonomously. The hyper-parameters can be configured, suiting system properties such as network and application workload.

\item \textbf{Trustworthiness.} A system is trustworthy if it provides accurate  data without  compromising with the data security~\cite{noor2013trust, 7163017}. In particular, it is more challenging to provide trustworthiness when the data is stored distributed, and replicated across a decentralized network of nodes. DEMon relies on a leaderless \emph{quorum-based consensus} (LQC) protocol to provide a trustworthy data retrieval mechanism. 

\item \textbf{Robustness.} A system is robust when it is able to operate despite the occurrence of one or more failures~\cite{brewer2000towards,van2003astrolabe}. DEMon provides robustness by using a decentralized architecture for its control and data plane. Stochastic information dissemination helps to balance the network load, and it avoids a single point of failure that may bring to complete data loss. Moreover, information about a node can be always retrieved, even if that node is currently unavailable since the information of all nodes is probabilistically stored in a decentralized fashion.

\item \textbf{Efficient communication and data retrieval.} The DEMon communication protocol optimizes network efficiency during data dissemination and retrieval. Its information dissemination Gossip protocol is designed to avoid unnecessary data transmission, and the querying LQC protocol provides a latency-sensitive response in run-time. 

\end{enumerate}


\begin{figure}[!ht]
\centering
\includegraphics[width=0.8\linewidth]{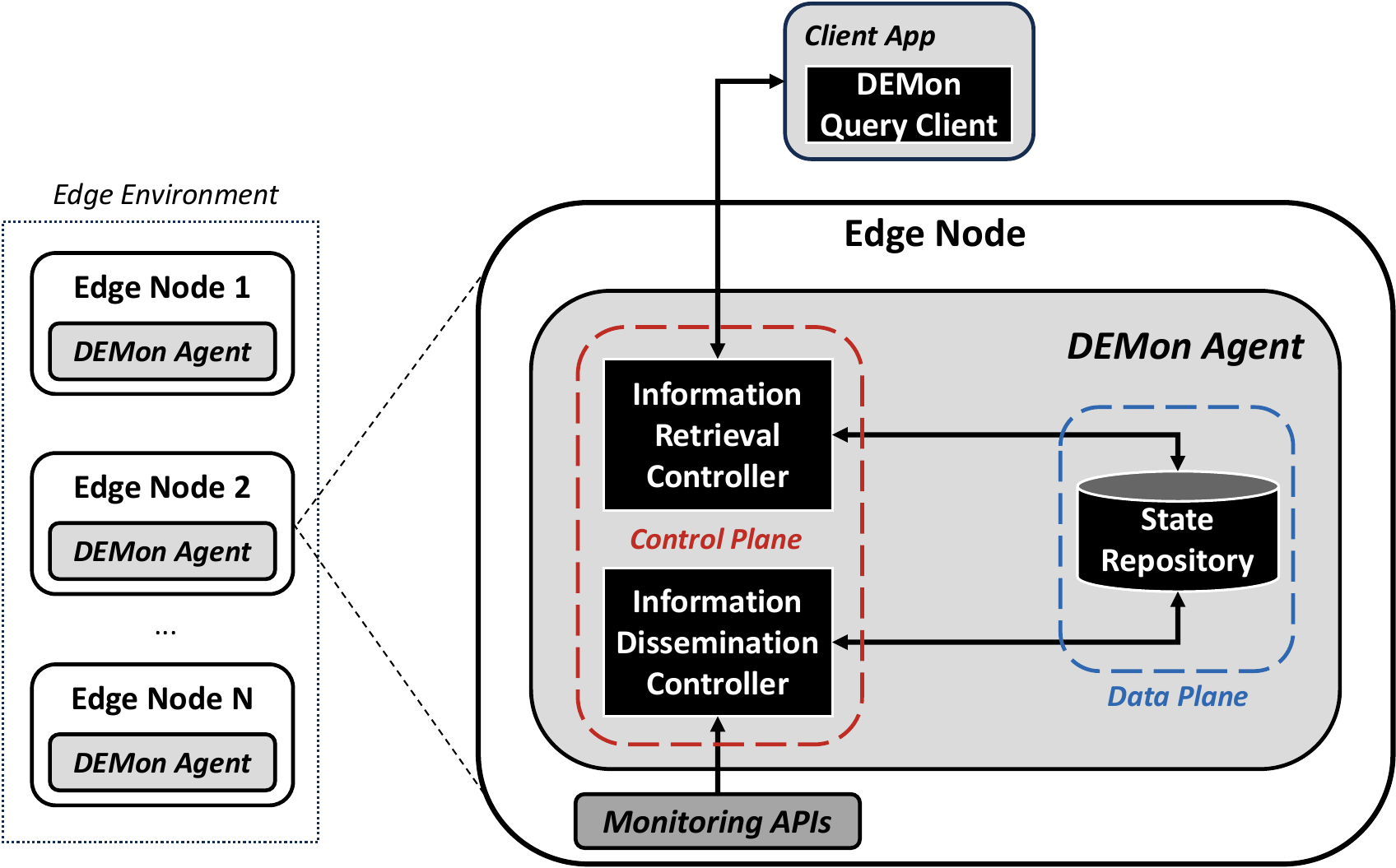}
\caption{\revisedmajor{}{The architecture of the DEMon monitoring system.}}
\label{fig:demon_arch} 
\end{figure}

Figure~\ref{fig:demon_arch} depicts the architecture of the DEMon monitoring system.
The system functionalities are realized within four main components:
\begin{enumerate*}[label=(\roman*)]
    \item State Repository,
    \item Information Dissemination Controller,
    \item Information Retrieval Controller, and
    \item DEMon Query Client.
\end{enumerate*}

The \emph{State Repository (SR)} contains the node states, that is, it contains both the monitoring data from the  node hosting the repository and monitoring data belonging to some or all of the other nodes participating in the network.

The \emph{Information Dissemination Controller (IDC)} is responsible for information dissemination, that is, it both sends and receives monitoring data leveraging a gossip-based algorithm. In particular, it collects monitoring data monitoring APIs (e.g., OS-specific or application APIs), and it stores them within the \emph{SR}. Then, it periodically sends the current system state, which includes its own state and the state of other nodes currently present in the \emph{SR}. Simultaneously, it waits asynchronously for the monitoring data sent by other nodes, and it only save the latest data to the \emph{SR}. 

The \emph{Information Retrieval Controller (IRC)} is responsible for information retrieval, that is, it provides an Application Program Interface (API) to retrieve monitoring data for any incoming request from client applications. It scans the \emph{SR} and returns back the requested information.

The \emph{DEMon Query Client (DQC)} is used by client applications that want to query the DEMon monitoring system and retrieve monitoring data. It provides a trustworthy data verification mechanism by leveraging the proposed Leaderless Quorum Consensus (LQC) protocol.

\revisedmajor{}{
Here, both the IDC and the IRC represent our control plane, and SR represents the data plane. Please note that each node running the DEMon Agent (DA) executes both controllers, and they are managed independently of each other. Each node also possesses its own SR, resulting in a homogeneous network of nodes with the same role and providing the same set of functionalities. Thus, our system provides a completely decentralized architecture for both the control plane and the data plane (SR). On the other hand, centralized monitoring systems like Prometheus~\cite{turnbull2018monitoring} or Zabbix~\cite{olups2010zabbix} possess a central controller and data plane, making them unsuitable for volatile edge environments. 
}
In the following subsections, we describe each of the DEMon components in detail.

\subsection{State Repository (SR)}
\label{subsec:state-repository}

The State Repository is a repository containing an historical record of node states. In particular, it both contains the states of the node hosting the repository, plus the states of some or all the nodes participating the network. Formally, $SR$ is a set of node states $\{S_{1}, \dots, S_{n}\}$, where $S_{i}$ is the set of states for node $i$ identified by an ID, e.g., its IP address. We define $S$ as a time ordered set of tuples $\{s_1, \dots, s_p\}$, where $s_t$ is the node state at time $t$. The node state $s$ is a tuple $(M, c, U, d)$, where $M$ is a set of metric values $\{m_1, \dots, m_k\}$ with $m_j$ representing the value of metric $j$ (e.g., CPU consumption); $c$ is an incremental counter that represents the number of gossip rounds; $U$ is the set node IDs that were unable to reach a node when selected for a gossip round; and $d$ is the resulting hash value (digest) obtained by computing the hash function $hash(M, c, U)$.

Listing ~\ref{lst:node-states} shows a concrete example of entry of the SR for a node identified by its IP address. The mapping with the tuple $S$ is straightforward: \texttt{metrics} represents the set of metrics $M$, \texttt{counter} represents the counter $c$, \texttt{unreachable\_by} represents the set of node IDs $U$, and \texttt{digest} represents the hash value $d$.
\\
\begin{lstlisting}[language=json,label={lst:node-states},numbers=none,tabsize=1,caption=Example of a State Repository entry.]
{
    "192.168.167.100": [
        {
            "metrics": {
                "cpu": 35,
                "memory": 2048,
                "storage": 10,
                "network_latency": 4,
                "heartbeat": 1695131028
            },
            "counter": 2,
            "unreachable_by": ["192.168.167.102"],
            "digest": "6ec3e0143db76fc0fbdd9d00311230fe4a1cfb565b73909fd9286a23f5e168a1" 
        },
        {
            ...
        }
    ]
}
\end{lstlisting}

\subsection{Information Dissemination Controller (IDC)}
\label{subsec:information-dissemination-controller}

The IDC is responsible for disseminating information (i.e., monitoring data) across the network of nodes leveraging a gossip-based protocol for an efficient data distribution. In particular, it provides two main functionalities: (i) it periodically collects monitoring data from its own execution environment (e.g., resource consumption or application-specific metrics), stores the collected values to the \emph{SR}, and disseminates the monitoring data to a subset of nodes participating the network; (ii) it asynchronously waits for incoming monitoring data from the other nodes, and it stores only the latest to the \emph{SR}. Each node can opportunistically configure the controller configuration parameters, i.e., time interval for data sending (\texttt{gossip\_rate}) and number of messages (\texttt{gossip\_count}), enabling control of the compute and network resource consumption at run-time, and information convergence speed.

To exemplify the two IDC functionalities, we illustrate the dissemination logic using the pseudo-code in Algorithms~\ref{alg:information-dissemination-controller-send} and~\ref{alg:information-dissemination-controller-receive}, respectively. From now on we use the term \emph{sending node} when referring to a node executing Algorithm~\ref{alg:information-dissemination-controller-send}, and \emph{receiving node} when referring to a node executing Algorithm~\ref{alg:information-dissemination-controller-receive}. However, please note that both the algorithms run in all the edge nodes, guiding the logic for sending and receiving the monitoring data of each node.

\begin{algorithm}[h]
\caption{Information Dissemination Controller - Send Node States}
\label{alg:information-dissemination-controller-send}
    \SetAlgoLined
    \For{every gossip\_rate}{
        
        $s \gets$ make\_new\_node\_state(); \Comment{Collect monitoring data and compute digest}

        $SR  \gets$ store\_node\_state($s$);

        $nodes$ $\gets$ select\_nodes\_to\_gossip();

        \For{$n \in nodes$}{
        
            \eIf{$|n.U| \geq failures\_threshold$}{
                $SR \gets$ delete\_node($n.id$);
            }{
                $SR\_metadata \gets$ get\_SR\_metadata(); \Comment{Node IDs and counters}

                $response \gets$ disseminate($n.id$, $s$, $SR\_metadata$);

                \eIf{$response$}{
                    $updates, requests \gets$ parse\_response($response$);
                    
                    \For{$u \in updates$}{
                        $SR \gets$ store\_node\_state($u$);
                    }
                    $requested\_node\_states \gets$ get\_requested\_node\_states($requests$);

                    send\_node\_states($n.id$, $requested\_node\_states$);
                }{
                    $SR \gets$ update\_unreachable\_node($n.U$);
                }
            }
        }
        
    }
\end{algorithm}

Algorithm~\ref{alg:information-dissemination-controller-send} shows the logic executed by a \emph{sending node}, which gossips monitoring data to the other nodes of the network.  Here, for every time interval defined by the \verb+gossip_rate+ parameter, the IDC first computes the new node state $s$ by gathering monitoring data (line 2), and it updates the $SR$ (line 3). Then, it \textit{randomly} chooses a subset of nodes equal to the configured \verb+gossip_count+ parameter from the $SR$ (line 4) to start gossiping. If a selected node has been already declared unreachable by a predefined configurable number of distinct nodes (i.e., \texttt{failures\_count}) (line 6), then such node is considered as failed and deleted by the $SR$ (line 7). More details about the deletion logic are provided in Section~\ref{subsubsec:node-network-management}. 

\revisedmajor{}{
Since a sender can choose random nodes to gossip, it is necessary to avoid duplicate or unnecessary data communication. If a receiver node already has the same or the most recent data as the other nodes, the sender should avoid gossiping such  data to the receiver. To address this, initially, the \emph{sending node} only sends its node state $s$ and metadata of its current known information stored in the $SR$ (line 9). Metadata includes all the known node IDs and their corresponding counters ($c$). This logic avoids excessive bandwidth consumption by excluding sending the duplicate data repetitively between the nodes. 
}
It is also important to note that, a \emph{receiving node} receiving the gossip message only intends to obtain missing or up-to-date data from other nodes. Consequently, the \emph{receiving node} replies back with a list of node IDs whose up-to-date state is requested (\texttt{requests}), and all the node states (\texttt{updates}) that it has fresher state $s$ than the sending node (i.e., because of the counters present in the metadata) (line 12). The \emph{sending node} then stores new node states within its $SR$, and also sends back the requested node states to the \emph{receiving node} (lines 13-17). Therefore, by choosing to only communicate the metadata first, and then sending the actual requested data, DEMon avoids excessive resource consumption of edge nodes. Finally, if there is no response, the \emph{receiving node} is marked as unreachable (line 19).

\begin{algorithm}[h]
\caption{Information Dissemination Controller - Receive Node States}
\label{alg:information-dissemination-controller-receive}
    \SetAlgoLined
    \While{$true$}{
        $sender\_node\_state, SR\_metadata \gets$ parse\_received\_message();
        
        $SR \gets$ store\_node\_state($sender\_node\_state$);

        \For{$node \in SR\_metadata$}{
        
            $current\_node\_counter \gets$ get\_node\_counter($node.id$);
            
            \eIf{$node.counter > current\_node\_counter$}{
                $requests \gets$ $node.id$;
            }{
                $updates \gets$ get\_node\_state($node.id$)
            }
        }

        send\_updates\_and\_requests($updates$, $requests$);
    }
\end{algorithm}

Algorithm~\ref{alg:information-dissemination-controller-receive} shows the logic executed by a \emph{receiving node}, which receives a gossip message from the other nodes in the network. For any incoming message, it first updates the $SR$ with the sender node state (line 3). Then, it checks if its $SR$ needs to be updated based on the metadata it has received, i.e., node IDs and counters (lines 4-10). Based on the comparison, the \emph{receiving node} sends back a response message composed by two parts: (i) node IDs requiring fresher states (\texttt{requests}), (ii) up-to-date node states that the \emph{sending node} does not have yet (\texttt{updates}). Thus,  the \emph{sending node} also receives new node states from the \emph{receiving node} within the same communication cycle.  In fact, the \emph{sending node} is simultaneously disseminating new information across the network, but also getting up-to-date data (if any) from the \emph{receiving node} in single gossip round. This two-way communication enables quick information sharing between nodes, enabling efficient and faster information dissemination. 

The stochastic selection of nodes in Algorithm~\ref{alg:information-dissemination-controller-send} ensures the uniform distribution of messages in the network, and it does not spike a specific network link, increasing  bandwidth usage. Moreover, the information spreads exponentially across the nodes and evenly distributes the network load~\cite{apt2017gossipcomputational}. Any permanent or transient failures will not affect the monitoring infrastructure. New nodes can send and receive monitoring data from the network and quickly know about all the nodes.

\subsubsection{Nodes Network Management}
\label{subsubsec:node-network-management}
Volatile edge environments are characterized by a dynamic network topology, with nodes joining and leaving at any time unpredictably, thus, enabling dynamic scalability for resource optimization ~\cite{offloadedge,bartolomeo2023oakestra}. To this extent, efficiently managing nodes' membership in an edge network is crucial, as the system size (i.e., the number of nodes) can change rapidly. However, this complicates the nodes' network management regarding new node insertion and deletion of a failed node in the monitoring system, leveraging a gossip-based communication protocol.

\paragraph{\textbf{Node Insertion}} DEMon assumes any node can join the edge monitoring network without any extra network configuration: they just need to know the identifier (e.g., IP address) of at least one of the nodes to start gossiping. In fact, when a new node with an unknown identifier joins the system and sends its state to a node already participating in the monitoring network, the latter inserts the new node state $s$ in its $SR$ according to the IDC logic explained above. This step allows the new node to participate in the gossip-based communication protocol, exchange data with other nodes, and contribute to the overall monitoring process.

\paragraph{\textbf{Node Deletion}} On the other hand, a node can also disconnect from the network for various reasons, such as resource or network failures, or even voluntary disjoining. Therefore, it is important to detect and delete a disconnected node from the $SR$ to reduce the data storage and increase the gossip-based protocol reliability (i.e., avoiding choosing a disconnected node for a gossip round). However, distinguishing between a node that is experiencing transient and temporary failures (e.g., short network disconnection), and a node that failed is a challenging problem ~\cite{aral2020learning}. If a node is deleted too aggressively, i.e., because of not receiving a single response, information re-convergence could take several rounds. Hence, a safe and reliable deletion is a key functionality of a decentralized monitoring system.

The DEMon embeds the node deletion strategy within the IDC logic (See Algorithm \ref{alg:information-dissemination-controller-send}). When a \emph{sending node} does not receive a response from a \emph{receiving node} during a gossip round, it marks such node as unreachable (i.e., it adds its own identifier $ID$ to the set $U$, denoting that it identified a potential failed node). This information is also gossiped in the following round as part of the node state $s$. When the size of the $U$ set of a node reaches the \texttt{failures\_threshold} (e.g., 3), that is, $k = failures\_threshold$ distinct nodes were unable to gossip to such node, DEMon considers it as failed or disconnected, and deletes the node from the $SR$. This ensures that DEMon does not delete any node for a transient failure or temporary network issues.

The node deletion strategy introduces reliability through distributed observation of failures and synchronization of this information across all nodes without any central coordination. This logic also affects communication and storage efficiency, as it stops the propagation and retention of such failed node data within the $SR$. By discontinuing the gossiping to a failed node and saving data storage, DEMon optimizes its resource allocation, ensuring that resources are not wasted on processing unreliable information. Furthermore,  the decisively failed status is communicated throughout the network as part of the usual IDC, in the subsequent gossip rounds, enabling other nodes to discover failure simultaneously.  

Nevertheless, when a failed node engages in gossiping and receives a valid response from another node, it signifies a restoration of connectivity, resulting in the reset of its decisive failed status. 
This adaptive mechanism acknowledges the importance of allowing nodes to recover from temporary failures and participate actively in the monitoring process, ensuring that network reliability is upheld. 
Please note that DEMon assumes the presence of security overlay mechanisms such as node authentication and authorization, which is out of the scope of this work.

\subsubsection{State Repository Management}
\label{subsubsec:state-management}

The $SR$ can be represented as a key-value data structure where the \texttt{key} of each entry is the node identifier $i$ (e.g., its IP address), and the \texttt{value} is the historical set of node states $S_i$. Such data structure can be persisted in memory, allowing fast read/write operations.
For an incoming gossip message, the IDC checks if it contains any new state for a node, comparing the current counter $c$ to the one received, and updating the $SR$ if needed. Thus, the $SR$ size (i.e., number of keys) remains constant at any time after the system converges unless new nodes join the system, allowing greater scalability.

However, the amount of memory required to store node states exponentially grows because at each gossip round a new node state $s$ is added to the node states set $S$, potentially creating memory issues on resource-constrained edge devices. To this extent, DEMon only stores the recent $r$ gossip rounds data of all nodes in-memory, and checkpoints at regular intervals to centralized data storage, enabling offline processing and analysis if required.
For example, in every $r$ elapsed gossip rounds, all the previous $r - 1$ node states are persisted to the centralized storage, and the in-memory $SR$ is refreshed retaining only the latest node states. The checkpoint interval can be either static or dynamically adaptable based on the gossip round, and each node can independently govern its own checkpoint interval. 

Checkpoints coming from different nodes have probabilistically duplicated node states, resulting in an exponential increase in the checkpointing data size. To further optimize the checkpointing logic to avoid duplicated storage, we store a single entry for each edge node in the checkpointing data storage, and only new data received from other nodes is additionally persisted (i.e., based on its counter values in $SR$). This ensures that the centralized storage maintains an up-to-date representation of the node states without excessive data redundancy.

These design choices optimize the utilization of resources within the edge environment. In fact, by keeping the most relevant information in memory, nodes can efficiently engage in real-time monitoring and data sharing. Periodic checkpointing ensures that a historical record of information is maintained, allowing comprehensive monitoring of the system designed for both online and offline batch processing.

\subsubsection{Illustration and Running Example of the IDC}

\begin{figure}[!ht]
\centering
\includegraphics[width=0.8\linewidth]{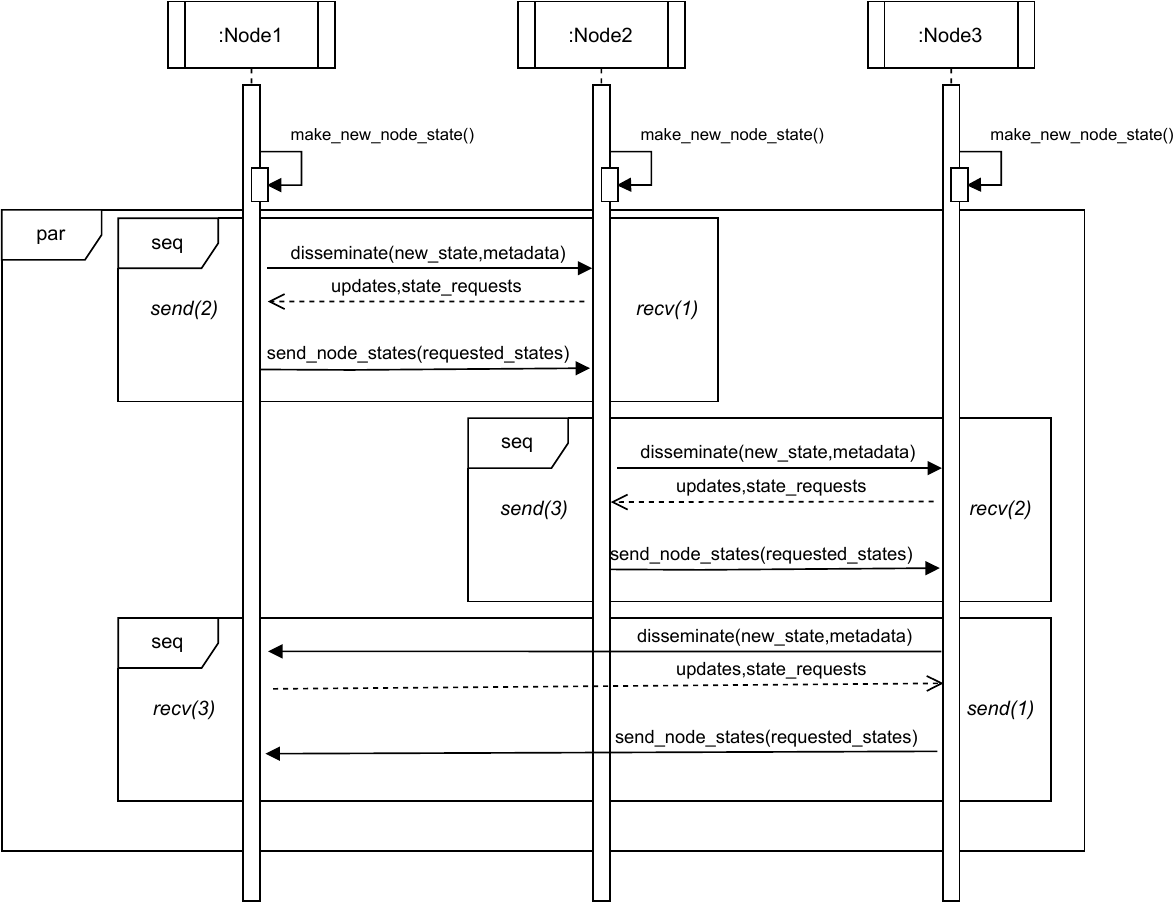}
\caption{
\revisedmajor{}{The sequence diagram of the DEMon monitoring system, illustrates the flow of messages between nodes in one round of gossiping.}
}
\label{fig:demon_seq_diagram} 
\end{figure}

\revisedmajor{}{
Figure \ref{fig:demon_seq_diagram} presents a sequence diagram that illustrates the sequence of messages exchanged among nodes during a gossip round as part of the Information Dissemination Controller (IDC). To simplify, we depict three nodes engaged in gossiping with \texttt{gossip\_count} set to 1, exchanging monitoring information. Although the diagram shows a sequential message exchange, it is important to note that all nodes gossip concurrently. For example, Node 1 initiates a gossip send message to Node 2 through random selection (\textit{send(2)}) with its current metadata (node IDs and counters) from $SR$. Upon receiving the message (\textit{recv(2)}), Node 2 compares the metadata with its $SR$ and sends back any updated data it has, as well as requests for any data missing from its $SR$. Subsequently, Node 1 sends the requested data to Node 2, completing Node 1's gossip cycle for that round. It is crucial to note that, in parallel, Node 1 is also receiving incoming gossip messages from other nodes during the same round. For instance, while Node 1 is communicating with Node 2, Node 3 sends a gossip message to Node 1, and Node 2 sends a gossip message to Node 3. This parallel communication accelerates information dissemination. The number of messages that are sent in a gossip round can be controlled with the \texttt{gossip\_count} parameter, further speeding up the information dissemination per gossip round. 
}

\begin{table}[h]
\caption {\revisedmajor{}{Illustration of state management with possible DEMon runtime operations (system size of 5). Notations R: Round; A: Action(s); SR: State Repository (with [$SR_1$] $\rightarrow$ [$SR_2$]: $SR_1$ $=$ begin of round \& $SR_2$ $=$ end of round); U: unreachable\_by (with (x,y) $\rightarrow$ z: z is unreachable by x \& y).}}
\label{table:demon-process} 
\centering
\setredmajor
\begin{tabular}{|c|c|c|c|c|c|c|} 
\hline
R                  &    & Node 1                      & Node 2                      & Node 3                             & Node 4                     & Node 5                       \\ 
\hline
\multirow{3}{*}{1} & A  & \textit{send(2), recv(3)}   & \textit{send(3), recv(1)}   & \textit{send(1), recv(2), recv(4)}          & \textit{send(3),accept(5)} & \textit{membership(4)}       \\
                   & SR & {[}1]$\rightarrow$[1:3]     & {[}2]$\rightarrow$[1:3]     & {[}3]$\rightarrow$[1:4]          & {[}4] $\rightarrow$[3:5]   & {[}5]$\rightarrow$[4,5]      \\
                   & U  & {[} ]                       & {[} ]                       & {[} ]                              & {[} ]                      & {[} ]                        \\ 
\hline
\multirow{3}{*}{2} & A  & \textit{send(3)}            & \textit{send(3)}            & \textit{send(4), recv(1), recv(3)} & \textbf{Internal Error}    & \textit{send(4)}             \\
                   & SR & {[}1:3]$\rightarrow$[1:4]   & {[}1:3]$\rightarrow$[1:4]   & {[}1:4]$\rightarrow$[1:4]        & -                          & ~[4,5]$\rightarrow$[4,5]                       \\
                   & U  & {[} ]                       & {[} ]                       & (3) $\rightarrow$ 4                & -                          & (5) $\rightarrow$ 4          \\ 
\hline
\multirow{3}{*}{3} & A  & \textit{send(5)}            & \textit{send(4), recv(3)}   & \textit{send(2), recv(5)}          & Internal Error             & \textit{send(3), recv(1)}    \\
                   & SR & {[}1:4]$\rightarrow$[1:5]   & {[}1:4]$\rightarrow$[1:4]   & {[}1:4]$\rightarrow$[1:5]          & -                          & {[}4,5]$\rightarrow$[1:5]    \\
                   & U  & (5)$\rightarrow$4           & (2,3) $\rightarrow$ 4       & (3,5) $\rightarrow$ 4              & -                          & (3,5) $\rightarrow$ 4        \\ 
\hline
\multirow{3}{*}{4} & A  & \textit{send(4), recv(2)}   & \textit{send(1), recv(5)}   & \textit{send(5)}                   & Internal Error             & \textit{send(2), recv(3)}    \\
                   & SR & {[}1:5]$\rightarrow$[1:3,5] & {[}1:4]$\rightarrow$[1:3,5] & {[}1:5]$\rightarrow$[1:5]               & -                          & {[}1:5]$\rightarrow$[1:3,5]  \\
                   & U  & (1:3,5) $\rightarrow$ 4     & (2,3,5) $\rightarrow$ 4     & (3,5) $\rightarrow$ 4              & -                          & (2,3,5) $\rightarrow$ 4      \\
\hline
\end{tabular}
\end{table}

We further illustrate the IDC process with an example depicted in Table \ref{table:demon-process}. This table provides a high-level overview of the state information $SR$, for each node at the start and end of a gossip round. For simplicity, we consider an edge system comprising 5 nodes with a \texttt{gossip\_count} of 1. We define a gossip round as $R$and actions as $A$. Within the state repository $SR$, we display only the node ID, represented as an integer. The list $U$, denoting \texttt{unreachable\_by}, contains the IDs of nodes unresponsive to gossip messages. It is important to note that $U \in SR$, but for clarity, we present it separately.   We consider that $SR$ \& $U$ is updated at the end of each gossip round and \textit{send()} \& \textit{recv()} only propagate the  $SR$ and $U$ with the updated state of the previous round.

Initially, Nodes 1-4 are part of the network, each possessing its own monitoring knowledge in their $SR$. In round $R1$, Node 5, not yet present in any  $SR$, announces its network membership by initiating the gossip process. It randomly selects Node 4 (\textit{send(4)}) and transmits a gossip message containing its metadata from  $SR$. Concurrently, Node 4 receives the message, acknowledges Node 5's membership, and incorporates Node 5's ID into its  $SR$.  Meanwhile, Nodes 1-3 continue to \textit{send()} and \textit{recv()} messages as  depicted in Figure \ref{fig:demon_seq_diagram}, with Node 4 also disseminating its $SR$ to Node 3.

During round $R2$, Node 4 fails, and Nodes 3 and 5 attempts to gossip (\textit{send(4)}) to it. The absence of a response leads them to add Node 4's ID to their respective $U$ lists, which will be propagated in the subsequent round.

In round  $R3$, Node 2 sends a gossip message to Node 4 and, receiving no reply, adds Node 4's ID to its  $U$. Simultaneously, Node 3 updates its  $SR$ after receiving a message from Node 5.

Finally, in round  $R4$, Node 1 attempts to communicate with Node 4 and, encountering no response, adds Node 4's ID to its   $U$. Node 1 also updates its  $SR$ based on a message from Node 2. At this stage, at Nodes 1, 2, and 5   $failures\_threshold$ reaches 3 (from list $U$) of Node 4, prompting the dissemination of its failed state across the network.  Consequently, Nodes 1, 2, and 5 remove Node 4's ID from their     
$SR$, as the  $U$ count for Node 4 hits 3. Additionally, the system achieves convergence, with all active nodes aware of each other. This process continues, ensuring that information regarding new and failure nodes is circulated autonomously as part of the gossiping monitoring information in the network.

\subsection{Information Retrieval Controller(IRC)}
\label{subsec:information-retrieval-controller}
The IRC provides an API to access the $SR$ and retrieve node states. Its functionality is straightforward, that is, it scans the $SR$ looking for the requested node IDs, and returns the corresponding node state sets ${S_1, \dots, S_n}$. Currently, the IRC does not provide any query language to filter or sort data, it only accesses node states by using the node identifier. However, a more sophisticated query language can be incorporated into the IRC to enable enhanced query capabilities.

\subsection{DEMon Query Client (DQC)}
\label{subsec:demon-query-client}
A critical aspect of DEMon is providing an efficient logic to obtain the monitoring information of an edge node in near real-time. Since data is replicated across the participant edge nodes, it is crucial to ensure that retrieved data is recent, consistent, and trustworthy. Generally, consistency is ensured in distributed information systems using the consensus protocols like Paxos~\cite{lamport2001paxos}. However, such protocols require the election of leader nodes, which creates a centralized bottleneck and increases latency, making them impractical in edge environments.

To this extent, we propose an efficient \emph{Leaderless Quorum Consensus (LQC)} protocol, where we eliminate the need for centralized leader nodes architecture. We introduce a leaderless consensus method where a client querying for the monitoring information reaches to consensus with decentralized participants. Such leaderless protocols have found applications in many recent distributed database systems such as Cassandra, 
\footnote{https://cwiki.apache.org/confluence/display/CASSANDRA/CEP-15\%3A+General+Purpose+Transactions}

allowing faster information retrieval. The protocol is implemented by DQC, which pseudo-code is shown in Algorithm~\ref{alg:query-client}.

\begin{algorithm}[ht]
\caption{DEMon Query Client implementing the Leaderless Quorum Consensus Protocol}
\label{alg:query-client}
    \SetAlgoLined
    \SetKw{KwGoTo}{go to}

    $query\_nodes \gets$ select\_random\_query\_nodes(quorum);\label{query-nodes}
	 
    $R \gets$ query\_metadata($query\_nodes$); \Comment{Node IDs and counters}
  
    
    \Comment{Provides a notion of weak consistency}
    
    \eIf{compare($R.timestamp$) is true}{

        \Comment{Ensures data trustworthiness}
        
        \eIf{ compare($R.digest$) is true}{
        
            \Return query\_data();
	}{
            \KwGoTo \ref{query-nodes};
        }
    }{ \KwGoTo \ref{query-nodes};}
	   
\end{algorithm} 

When the DQC has to perform a query to the DEMon system, it spawns parallel queries to a random subset of nodes (i.e., predefined \texttt{quorum}), and obtains responses consisting of the queried data, and the associated message digests (lines 1-3) from the IRC. Once it receives the minimum number of responses (i.e., \texttt{quorum}), it initially compares their counters (line 4). If the counter of responses matches, then corresponding digests are matched. If this condition holds, then the data is considered consistent and trustworthy. Otherwise, the current request session is discarded, and a new set of nodes is chosen randomly (lines 5-11).

Please note that the QC does not provide any feature for retrieving granular monitoring information because it just relies on the IRC API to obtain the node states. Rather, it provides the ability to retrieve data of a node or aggregated data of all nodes in a decentralized manner.

This protocol may not guarantee strong consistency (i.e., returning always the most recent data of a node), but it ensures a significant performance advantage, providing weak consistency. In addition, it provides the reliability required for a volatile edge environment by avoiding a single point-of-failure. Nevertheless, the deterministic and irreversible digest encoded in the node state ensures data integrity (monitoring information is not altered by a malicious node) to any query result. 

Trustworthiness can be even further improved by including mechanisms such as digital signatures, certificates, reputation management systems, and proof-of-work systems. However, such techniques significantly increase the computational and storage cost, thus, they are not advisable considering the resource constraints that characterize edge environments.

\section{Prototype Implementation}
\label{sec:implementation}

We implemented the DEMon monitoring system in a prototype available at \repo. Both the \emph{DEMon Agent (DA)} components and the \emph{DEMon Query Client (DQC)} are implemented using Python language.

The \emph{DA} exposes HTTP JSON APIs leveraging the Flask framework\footnote{\url{https://flask.palletsprojects.com/en/2.2.x/}} to implement the \emph{IDC} and \emph{IRC} functionalities. In particular, the \emph{IDC} registers an HTTP endpoint to handle incoming gossip messages, while the \emph{IRC} registers an HTTP endpoint to handle queries performed by the \emph{DQC}. Moreover, the \emph{IDC} spawns a dedicated thread to collect monitoring data (i.e., CPU and memory consumption, disk usage, and network I/O) using the \emph{psutil tool}\footnote{\url{https://pypi.org/project/psutil/}} and send gossip messages. The \emph{SR} is implemented as an in-memory Python dictionary, while the checkpoint's database is realized by an SQLite\footnote{\url{https://www.sqlite.org/index.html}} instance.

The \emph{DQC} is provided by a Python module that can be imported into third-party client applications that want to retrieve trustworthy monitoring data using the LQC protocol.

Finally, we provide a Dockerfile to build a Docker image and start the DEMon agent as a containerized application.


\section{Performance Evaluation}
\label{sec:evaluation}

This section describes how we conducted the performance evaluation and discusses the experimental results. In particular, we investigate the following four Research Questions (RQs) to evaluate our approach.

\begin{enumerate}[label=RQ$_\arabic*$:]
\item \textbf{Convergence. \textit{How quickly can DEMon discover all the nodes in the edge network and how do different parameters affect the convergence speed?}}   This RQ investigates how quickly each node knows about every other node participating in the network (i.e., convergence). Also measures the time required for a new node to know about the other nodes when it joins the network. It is important to note that each node will update its monitoring information continuously, and the gossiping continues forever to spread the recent information across the system. 
\item \textbf{Storage and Network. \textit{What is the storage and network overhead of the DEMon in the information dissemination process?}} This RQ investigates the overhead introduced by DEMon in terms of of storage and communication, and it studies whether the system could scale with different numbers of nodes on the edge.
\item \textbf{Query Latency. \textit{How many messages does it take to find the requested information using the proposed LQC protocol?}} This RQ investigates the number of messages required to retrieve the information about any particular node based on Algorithm \ref{alg:query-client}.
\item \textbf{Age of Information (AoI). \textit{How do we measure the quality of the information retrieved from  DEMon?}} Since the monitoring information is continuously updated, it is essential to know the timeliness of the information stored in each node. In our case, it is the freshness of the data in each node. To this end, we measure the Age of Information (\verb+AoI+) metric \cite{arafa2019timely}. The \verb+AoI+ of any node in the system is defined as:
    
    \begin{equation}
  \label{eq:aoi}
        \begin{aligned}
       AoI= \frac{1}{n}\sum_{i=1}^{n}   t_i - u(t_i)
        \end{aligned}
\end{equation}
    
where $n$ is the total number of other nodes' information a node has, $t_i$ is the actual time counter of remote $node_i$, and  $u(t_i)$ is the current time counter of $node_i$ locally.  To understand the \verb+AoI+,  let us assume the system has two nodes, \verb+node_1+, and \verb+node_2+, and the monitoring system has reached an initial convergence state (every node knows everyone). The \verb+AoI+  of   \verb+node_1+ is the difference between the time counter for \verb+node_2+ at  \verb+node_1+ i.e., $u (t)$ and the actual time counter at  \verb+node_2+ i.e., $t$. In our case, the time counter is an integer value (as described in  Section \ref{subsec:state-repository}); each node increments its time counter in a predefined time interval.
\end{enumerate}

\subsection{Baseline}
We select FogMon2~\cite{gaglianese2023assessing_fogmon2, fogmon} as the baseline to compare our approach. FogMon2 is a hierarchical peer-to-peer (P2P) monitoring system for Fog-Edge environments. Despite there are other relevant edge monitoring solutions available in the literature, such as  PyMon~\cite{grogman_2017} or FMonE~\cite{brandon2018fmone}, FogMon2 is the most suitable for a comparison with a fully decentralized system as DEMon.

FogMon2 leverages a hierarchical two-layer architecture composed of Leader and Follower nodes. Followers forward monitoring data to their Leader at regular intervals (i.e., report time), while Leaders share monitoring data with each other in a P2P fashion that is similar to a gossip-based algorithm. Among many configuration parameters available, the number of Leaders with respect to the size of the overall system is the most influential factor for the performance and robustness of FogMon2.

\revisedmajor{}{
We executed FogMon2 using Docker containers. On our testbed server, we tested with a system size of up to 80 nodes, which was the maximum possible on our testbed server. 

In particular, we experiment with multiple leader node sizes (i.e.,  \(2\times \sqrt{N}\), \(\sqrt{N}\), and \(\frac{\sqrt{N}}{2}\)), where $n$ is the total number of nodes in the system.  The number of Follower nodes is initially evenly distributed among all the Leaders and then balanced by FogMon2  itself.
}

\subsection{Experimental Setup}

We evaluate DEMon on a Docker-based testbed to create a large-scale realistic edge environment, feasible for repetitive and inexpensive experiments.
We execute Docker containers on a bare-metal server (40-core Intel(R) Xeon(R) CPU E5-2630L, 128GB RAM) hosted in our HPC laboratory. Each container executes the DEMon Agent and represents an edge node in our setup.
We scaled edge nodes (i.e., containers) from 0 to 300, with an interval of 50. 
Please note that no changes are required to deploy our monitoring system on the new edge infrastructure, either physical or virtualized, provided the nodes can run as Docker containers.

\begin{table}[]
\caption{The  hyper-parameters used in the experiments, their definition and configuration values}
\label{table:parameter}
\begin{tabular}{lll}
\hline
\multicolumn{1}{c}{\textbf{Parameter}} & \multicolumn{1}{c}{\textbf{Defintion}}                                                                      & \multicolumn{1}{c}{\textbf{Values}}    \\
\hline
gossip\_count (integer)                & \begin{tabular}[c]{@{}l@{}}Defines how many target nodes \\ are selected for each gossip round\end{tabular} & \{2, 3, 4\}                            \\
gossip\_rate  (seconds)                & \begin{tabular}[c]{@{}l@{}}Defines  the time interval between\\ two  gossip rounds\end{tabular}             & \{1, 5, 10, 15, 20\}                   \\
failure\_rate  (\%)                    & Defines \% of failed nodes                                                                                  & \{10, 20, 30, 40,  50 60, 70, 80, 90\} \\
n: system size (integer)               & \begin{tabular}[c]{@{}l@{}}Defines number of nodes in the\\ edge environment\end{tabular}                   & \{50, 100, 150, 200, 250, 300\}    \\
\hline
\end{tabular}
\end{table}

Our monitoring framework's performance depends on multiple hyper-parameters including \verb+gossip_count+, \verb+gossip_rate+, \verb+failure_rate+, and the system size $n$. Thus, we conducted experiments with different hyperparameter values and analyzed the results. The details of these hyper-parameters are presented in the Table \ref{table:parameter}.

All the experimental material to reproduce the experiments is available in our publicly released repository\footnote{\repo}.

\subsection{Results and Analysis}
\subsubsection{$RQ_1$: Convergence}

\begin{figure*}
\centering
\label{fig:main_results}
  \captionsetup{justification=centering}
   \begin{subfigure}[t]{0.49\textwidth}
        \includegraphics[width=\linewidth]{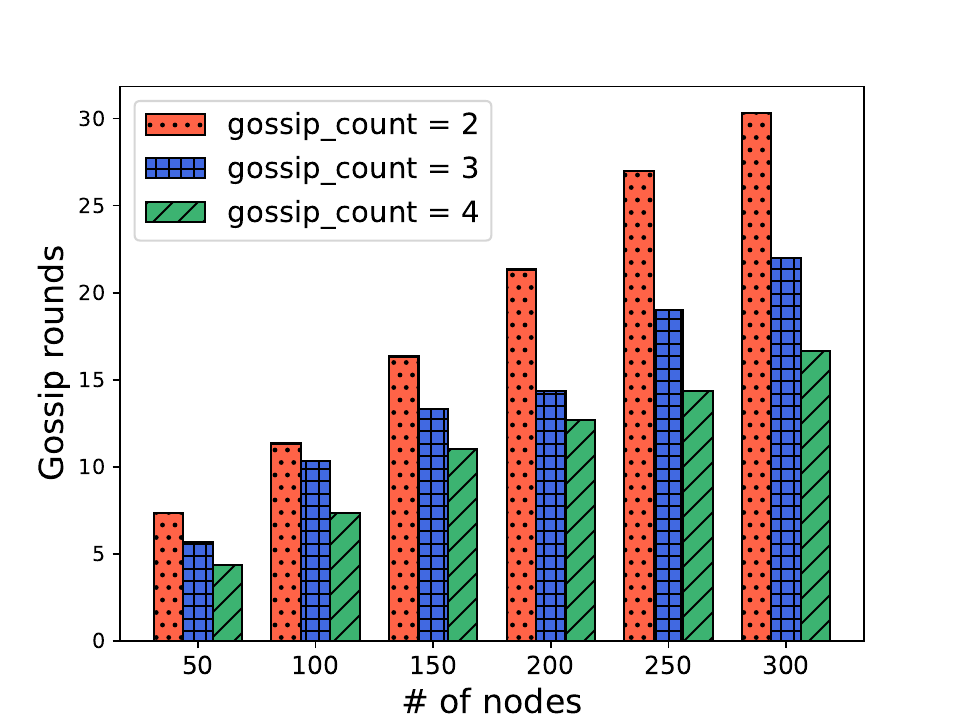} 
       \caption{Convergence vs rounds (gossip rate = 1).}
        \label{fig:rounds_converge}
    \end{subfigure}
    \begin{subfigure}[t]{0.49\textwidth}
        \includegraphics[width=\linewidth]{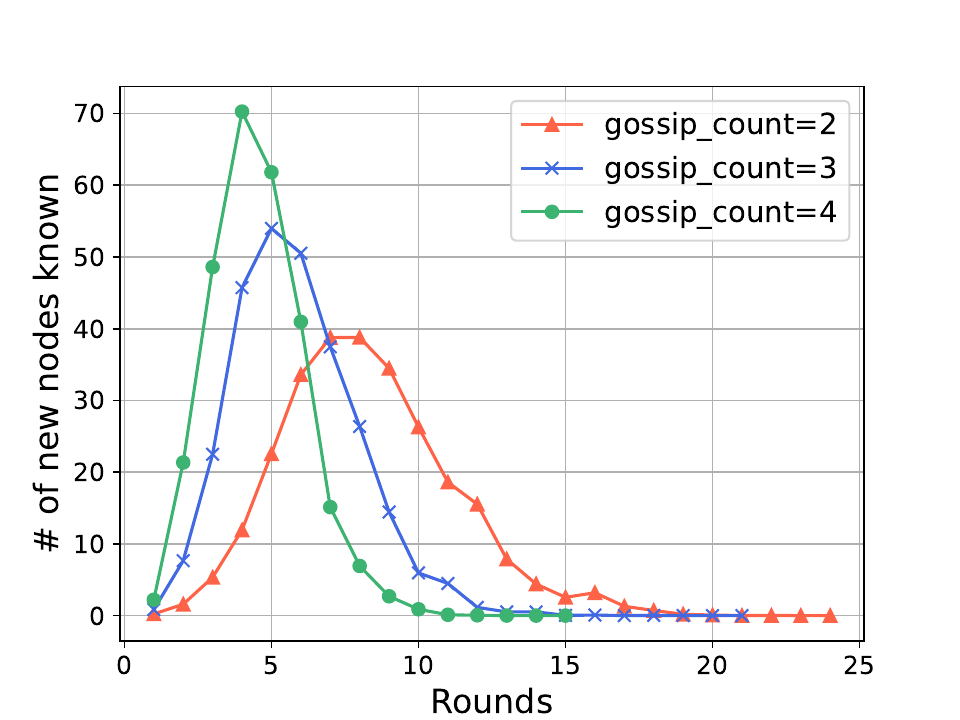}
        \caption{ Gossip rounds vs known nodes (system size=300,  gossip rate=1).  The higher the gossip count, the faster the information spreads. }
        \label{fig:rounds_knownnode}
   \end{subfigure}
   \caption{ Performance of gossip protocol. }
\end{figure*}

In the following experiments, we scaled the number of edge nodes, $n$ (i.e., containers) up to 300 with an interval of 50. All experiments are repeated three times, and average values are considered for the analysis.

Figure \ref{fig:rounds_converge} shows an analysis of the number of gossip rounds required to reach the initial convergence state. Since each node has its gossip round running locally, at any instant, we consider the maximum gossip round among all nodes as system convergence (when all nodes know about all other nodes). Here, we configured  \verb+gossip_count+ $ = \{2,3,4\}$  and  \verb+gossip_rate+ to 1. 
The higher value of \verb+gossip_count+ spreads the information quickly, and the system converges faster in lesser gossip rounds as observed in Figure \ref{fig:rounds_converge}. This behavior denotes that when a new node (re)joins the system,  it is able to know about all other nodes in fewer gossip rounds. For instance,  when gossip  \verb+gossip_count+ is set to 4, the system converges within four rounds ($n=50$ ).  Similarly, Figure \ref{fig:rounds_knownnode} shows the number of new nodes known (average value of all nodes) during each gossip round. As observed, more than 50\% of nodes' data is gathered within the first four gossip rounds ( \verb+gossip_count+ $=4$). 
\\
\textbf{Comparison with Baseline:}

    \begin{figure}[t]
        \includegraphics[width=\linewidth]{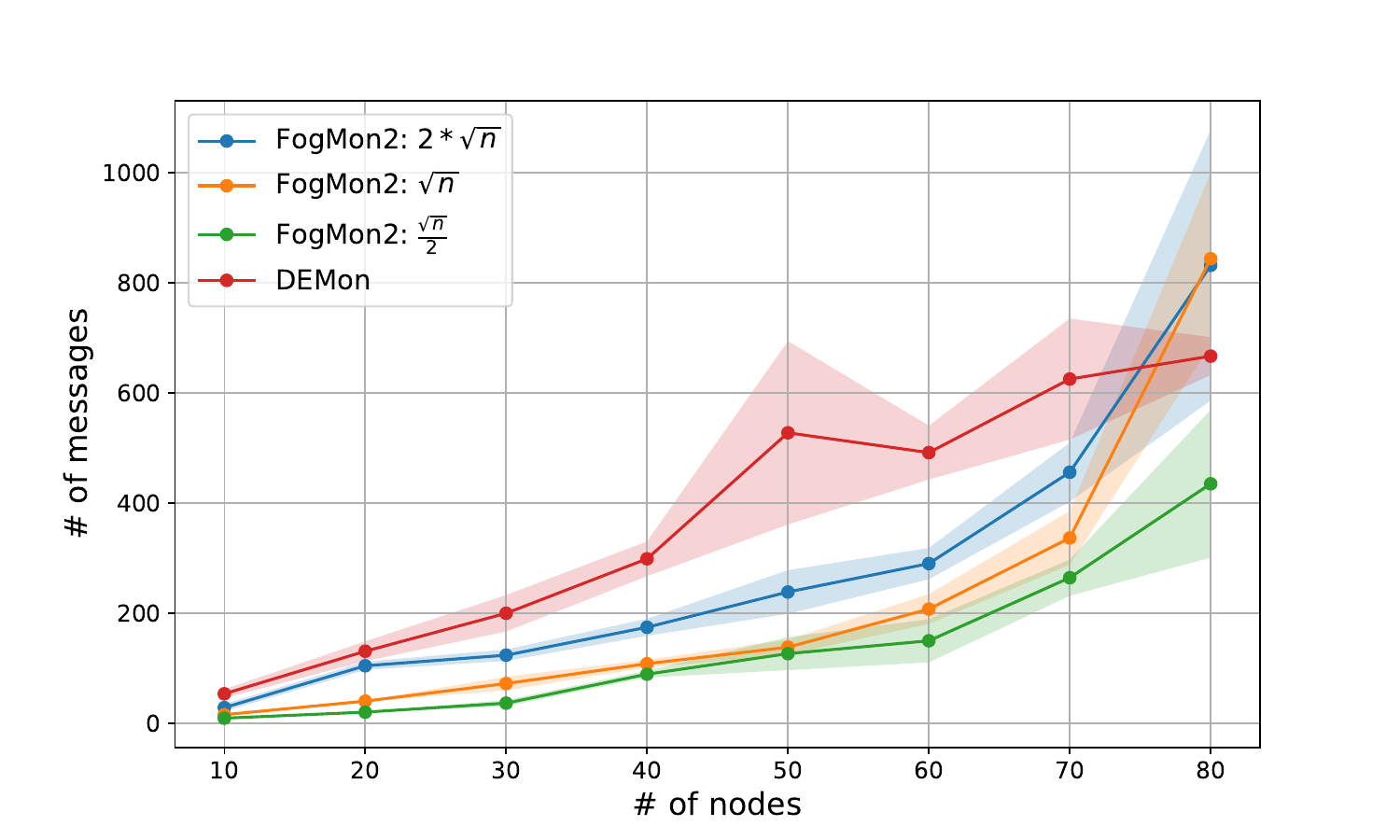}

   \caption{   \revisedmajor{}{ 
       A number of messages for convergence between FogMon2 and DEMon. Here, we set 3 different configurations of FogMon2 (number of Leader nodes) and gossip\_count=2 for DEMon.}
   }
    \label{fig:baseline_fogmon}
\end{figure}

\revisedmajor{}{ 
Figure~\ref{fig:baseline_fogmon} presents the performance of DEMon in comparison to FogMon2~\cite{gaglianese2023assessing_fogmon2}.
We modified the existing FogMon2 \footnote{\href{https://github.com/hpc-tuwien/DEMon/}{https://github.com/hpc-tuwien/DEMon/}} tool to collect the number of messages the Leader nodes send and receive until convergence is achieved.

The efficiency and convergence speed of FogMon2 can be influenced by the number of Leader nodes, which disseminate the monitoring information among themselves. We consider the system to be converged when all Leader nodes are aware of all participating nodes in the network. We tested three Leader node configurations as recommended by FogMon2: \(2\times \sqrt{N}\), \(\sqrt{N}\), and \(\frac{\sqrt{N}}{2}\). We used the number of messages required for convergence as a performance indicator.

As depicted in Figure \ref{fig:baseline_fogmon}, DEMon initially requires more messages due to its completely decentralized communication architecture, unlike FogMon2. However, as the system size grows, the gap in the number of messages between the optimal configuration of FogMon2 \(\sqrt{N}\) and DEMon narrows.  With a larger system size, DEMon and FogMon2 have comparable performance in terms of number of messages required for convergence, demonstrating DEMon's capability of being a completely decentralized system and performing similarly to partially decentralized approaches like FogMon2.
}


\begin{figure*}
\centering
  \captionsetup{justification=centering}
    \begin{subfigure}[t]{0.45\textwidth}
        \includegraphics[width=\linewidth]{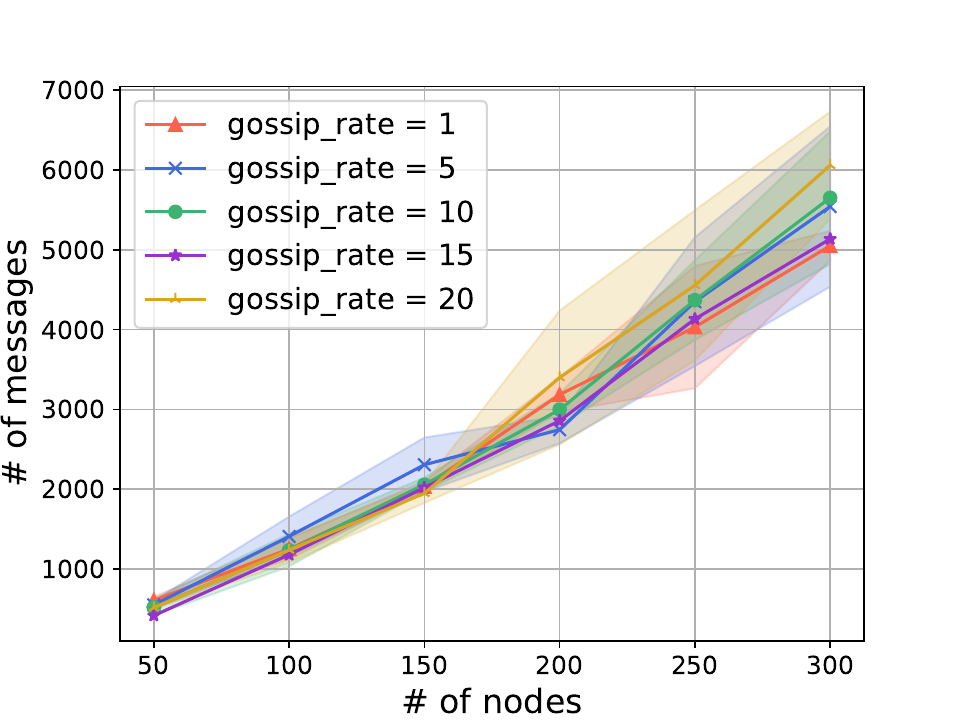}
        \caption{Effect of gossip rate on number of messages (gossip count = 3).}
        \label{fig:gr_messages}
    \end{subfigure}
      \begin{subfigure}[t]{0.45\textwidth}
        \includegraphics[width=\linewidth]{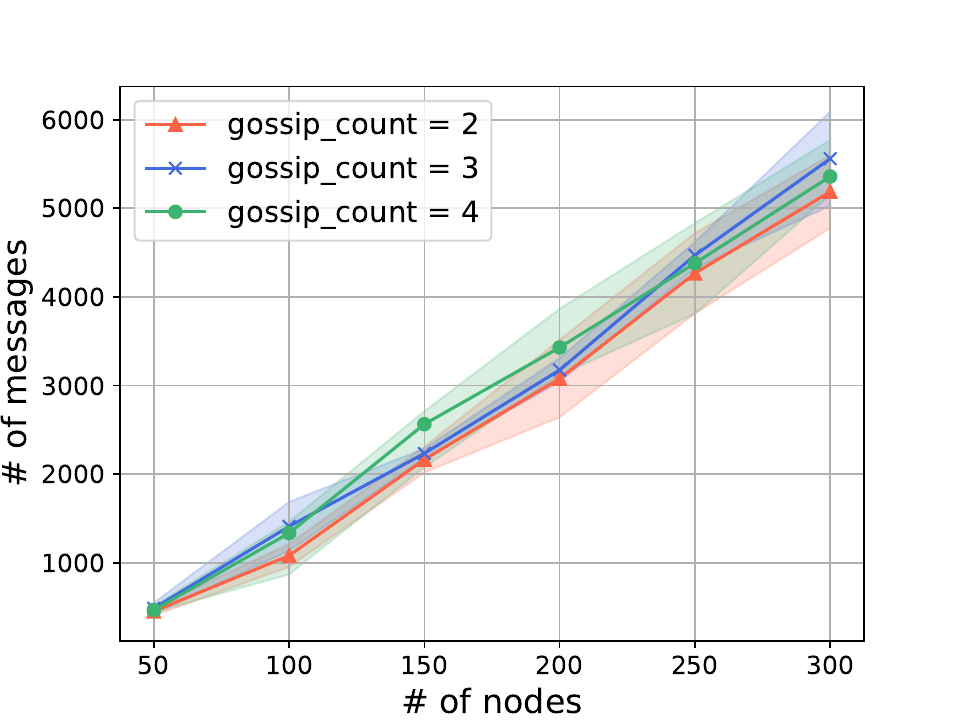}
        \caption{Effect of gossip count on number of messages (gossip rate = 1).}
        \label{fig:gc_messages}
    \end{subfigure}
    
 \caption{
    \revisedmajor{}{ 
    Sensitivity analysis: Effect of gossip rate and gossip count parameter  on the number of messages for  convergence  with different numbers of nodes.}
   }
    \label{fig:gr_gc_messages_plots}
\end{figure*}

\textbf{Hyper Parameter Analysis.} 
    Figure \ref{fig:gr_gc_messages_plots} depicts the effects of  \verb+gossip_rate+ and  \verb+gossip_count+ on the number of messages during the initial system convergence. Figure \ref{fig:gr_messages}  shows the number of messages it takes to converge the system with different \verb+gossip_rate+.  Here, the plots also depict the average value from three different runs and their standard deviations.
     Each node's monitoring interval and \verb+gossip_rate+ are set to similar; thus, for each \verb+gossip_rate+ second, a node sends its new state. We configured \verb+gossip_rate+ $=$ $ \{1,5,10, 15, 20 \} $ seconds and \verb+gossip_count+  to 3.   When the number of nodes increases, the total number of messages simultaneously increases, demonstrating that a higher number of message exchanges are required for many nodes, as shown in Figure  \ref{fig:gr_messages}. Irrespective of \verb+gossip_rate+, the number of messages is similar across all node sizes; representing convergence requires an almost equal number of messages for fixed system size and does not depend on \verb+gossip_rate+. 

Similarly, \verb+gossip_count+ also significantly affects the performance of the monitoring system. The  \verb+gossip_count+  decides how many random nodes are chosen for each gossip round.   Figure \ref{fig:gc_messages} depicts the effects of   \verb+gossip_count+ on the number of messages. We configured \verb+gossip_count+ $ = \{2,3,4\}$  and \verb+gossip_rate+  is set to 3 in these experiments. Similar to  \verb+gossip_rate+, number of messages do not vary significantly across different system size (see Figure \ref{fig:gc_messages}).  

\begin{figure*}
   \captionsetup{justification=centering}
        \centering
        \includegraphics[width=0.7\linewidth]{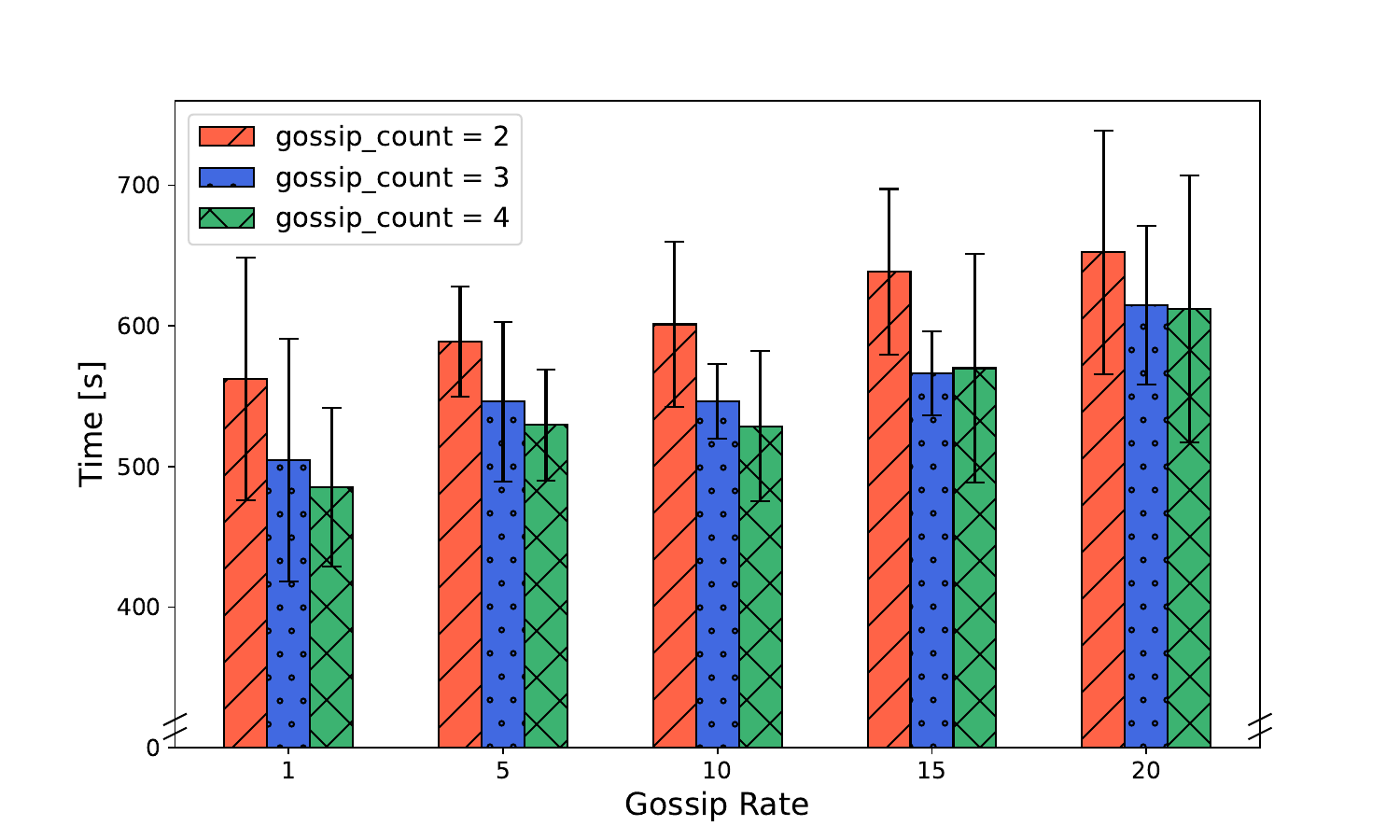} 

    \caption{
        \revisedmajor{}{Sensitivity analysis: Effect of gossip rate and gossip count parameter on the time for  convergence with different number of nodes (n = 300).
        }
    }
    \label{fig:gr_gc_time-plot}
\end{figure*}

\revisedmajor{}{
Although \texttt{gossip\_rate}  and  \texttt{gossip\_count} do not affect the number of messages it takes to converge the system, they significantly affect the convergence time. Figure \ref{fig:gr_gc_time-plot} shows the effect of these parameters on time. The x-axis in Figure \ref{fig:gr_gc_time-plot} shows different \texttt{gossip\_rate}, and grouped bar plots represent \texttt{gossip\_count}  configurations. As observed, the   \texttt{gossip\_rate} directly affects the convergence time since it controls the number of messages per second.
}

Therefore, if we want to control the bandwidth usage or speed of convergence, \verb+gossip_rate+ can be configured accordingly.  The smaller \verb+gossip_rate+ results in faster convergence time and vice versa. Similar observations can be drawn for the  \verb+gossip_count+ parameter. Thus, if faster convergence and a higher rate of information exchange are required,  \verb+gossip_count+ should be set to a higher value, and \verb+gossip_rate+ should be set to a smaller value. If minimal bandwidth consumption is necessary, contrast values can be set to these parameters.

\revisedmajor{}{
\textbf{Answer to RQ1:} DEMon significantly enhances the convergence rate of monitoring information (e.g., less than 25 rounds when n=300, \texttt{gossip\_count} =3), offering increased control over both the speed of convergence and resource utilization (number of messages). This is achieved through the fine-tuning of configuration parameters—specifically \texttt{gossip\_rate} and \texttt{gossip\_count}, which can be adjusted to meet the specific demands of the edge environment.
}

\subsubsection{ $RQ_2$: Resource Usage Analysis}

\begin{figure*}
  \captionsetup{justification=centering}
    \begin{subfigure}[t]{0.33\textwidth}
        \includegraphics[width=\linewidth]{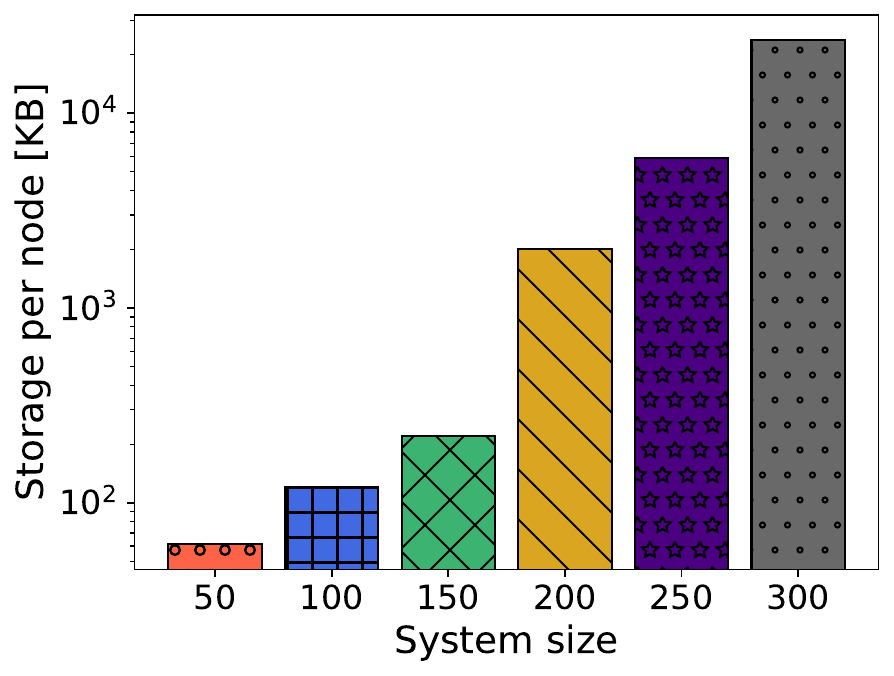} 
         \caption{Storage consumption after convergence (gossip\_count = 3).}
        \label{fig:storage_converged}
    \end{subfigure}
    \begin{subfigure}[t]{0.33\textwidth}
        \includegraphics[width=\linewidth]{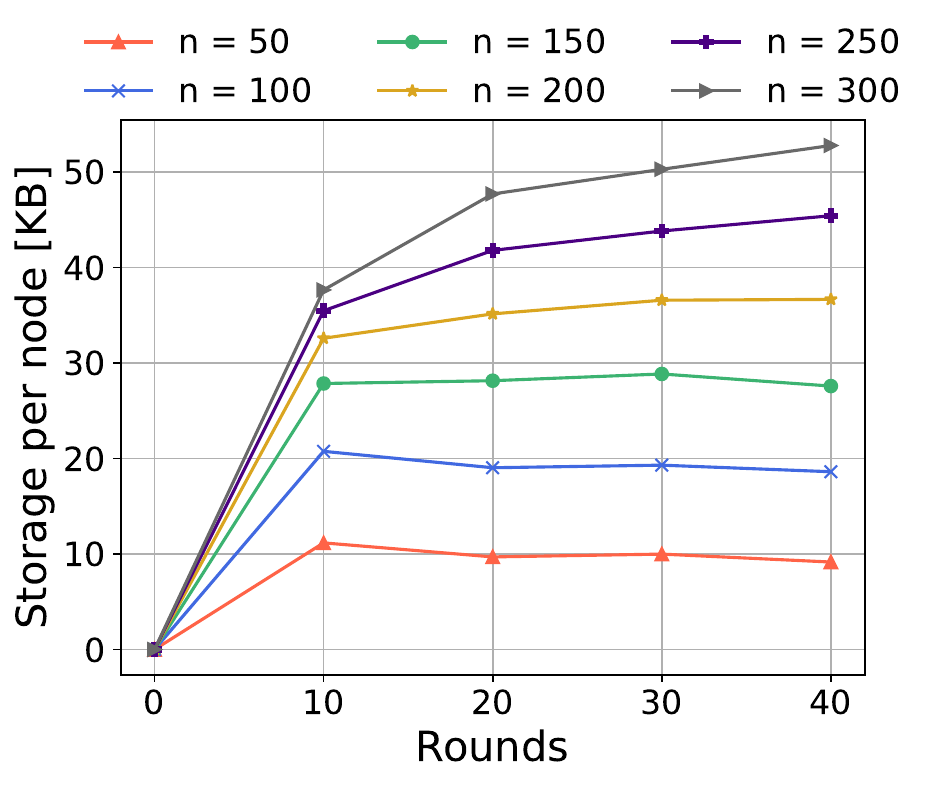} 
        \caption{Storage consumption every 10th round (gossip\_count = 3).}
        \label{fig:push_storage}
    \end{subfigure}
    \begin{subfigure}[t]{0.33\textwidth}
        \includegraphics[width=\linewidth]{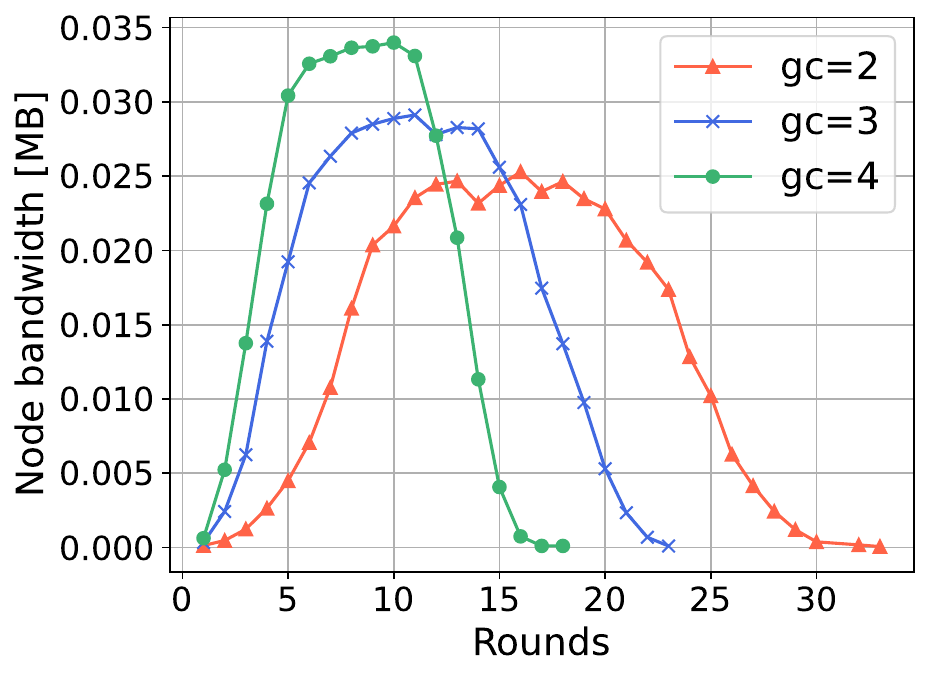} 
        \caption{Bandwidth till convergence (n = 300, gc = gossip\_count).}
        \label{fig:bandwidth}
    \end{subfigure}
    \caption{Analysis total storage after convergence, storage consumption over gossip rounds and bandwidth consumption over gossip rounds till convergence (gossip rate = 1).}
    \label{fig:storage_network}
\end{figure*}

\begin{figure*}
  \captionsetup{justification=centering}
  \begin{subfigure}[t]{0.49\textwidth}
        \includegraphics[width=\linewidth]{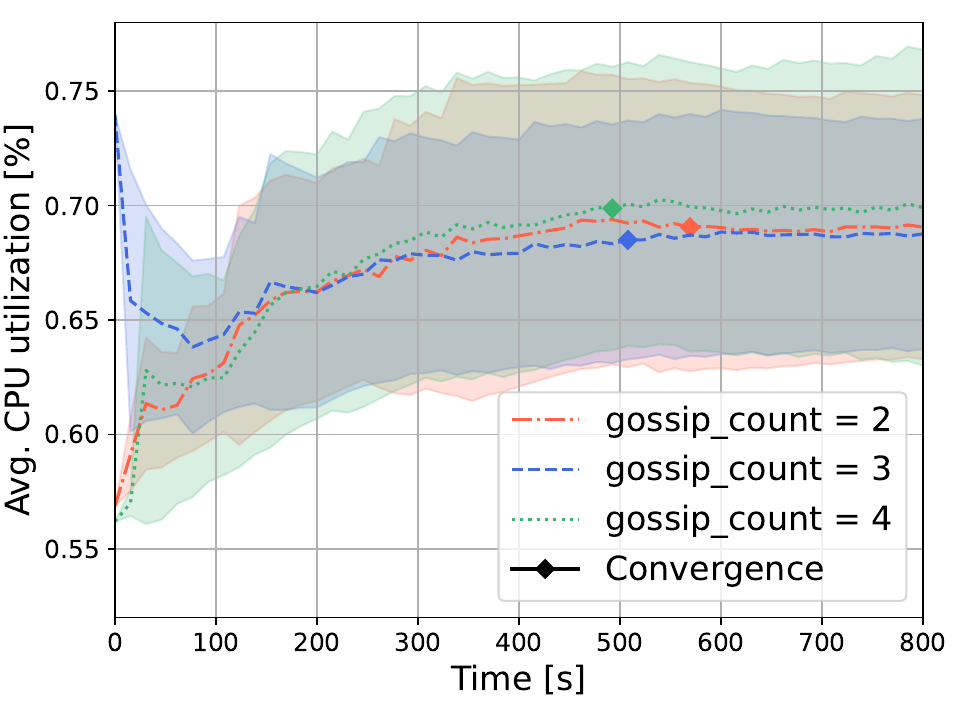} 
       \caption{
         \revisedmajor{}{CPU Usage}
         }
        \label{fig:cpuusage}
    \end{subfigure}
    \begin{subfigure}[t]{0.49\textwidth}
        \includegraphics[width=\linewidth]{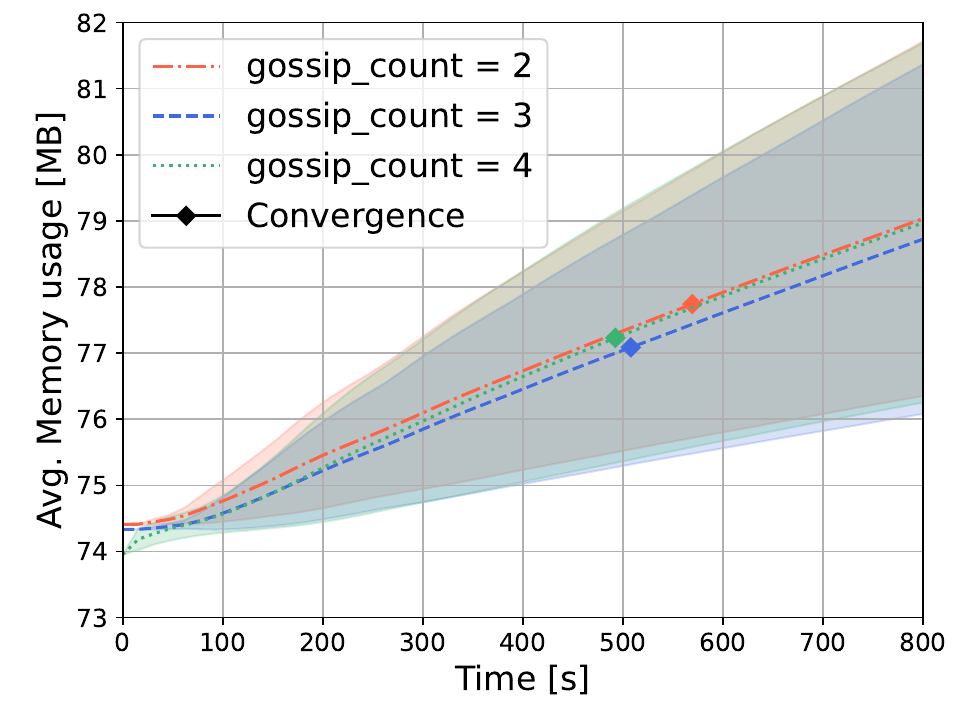} 
        \caption {
        \revisedmajor{}{
        Memory Usage  }
        }
        \label{fig:memusage}
    \end{subfigure}
   \caption{ 
   \revisedmajor{}{Analysis of Resource Usage by DEMon (system size=300, gossip\_rate=1). }
   }
    \label{fig:resource_usage_analysis}
\end{figure*}

To evaluate the storage and network overhead of DEMon, we measure the size of the state repository $SR$ at each node until convergence and after convergence, and also the bandwidth consumption across gossip rounds.

The average results across all nodes are presented in Figure \ref{fig:storage_network}. Figure \ref{fig:storage_converged} shows the average size of $SR$ with different system sizes when the overall system has converged. For instance, with a maximum system size of 300, the state repository has 23.68 MB (23698 KB)  in its size, demonstrating the lightweight memory footprint of DEMon. Similarly, Figure \ref{fig:push_storage} shows the size of $SR$ across different gossip rounds. The storage size increases exponentially until the system converges since new nodes' states are added to $SR$. After that, we have a linear increase and almost a constant size in storage size. It is important to note that, we refresh the in-memory $SR$ for every 10th round, to keep the most recent run-time data in the system. The average network bandwidth across different gossip rounds can be seen in Figure \ref{fig:bandwidth}. The network bandwidth has a similar trend as storage: it is high until convergence since a large amount of data needs to be gossiped initially. After the system converges, DEMon consumes extremely low levels of bandwidth. These empirical results of storage and network consumption show that DEMon can scale with low resource footprints on the edge resources.

We also analyze the CPU and memory resource consumption of DEMon agents. Figure \ref{fig:resource_usage_analysis} illustrates the CPU usage (Figure \ref{fig:cpuusage}) and memory usage (Figure \ref{fig:memusage}) of a DEMon agent. For the CPU usage analysis, we maintained a system size of 300, with a gossip rate set to 1. The figure presents the average resource consumption across all 300 edge nodes (containers). For each data point, we calculate the cumulative average and display the cumulative standard deviation. The marked points on the scatter plot indicate the time at which the monitoring system has converged (when each node knows every other node). As observed, our monitoring agent consumes less than 0.75\% of CPU resources, demonstrating DEMon's lightweight design and effectiveness in resource-constrained environments. The CPU consumption exhibits a consistent pattern even post-convergence, indicating stable CPU resource utilization.

Similarly, Figure \ref{fig:memusage} depicts the cumulative average memory consumption of all the edge nodes. Notably, until convergence, the memory consumption increases as new nodes' data is incorporated into \textit{SR}. Post-convergence, the memory consumption shows a decline. This behavior can be modulated based on the number of timestamps of past data retained in our system (currently, we store all historical data in memory to show the maximum possible usage). In conclusion, DEMon's resource utilization is suitable for resource-constrained edge environments due to its minimal CPU and memory footprints.

Furthermore, we examined FogMon2's CPU and memory consumption with a system size of 80 (the maximum possible on our server) and found that FogMon2, on average, exhibits a CPU usage of 5.9343\% and a memory usage of 42,217 MB across leaders and followers. The higher CPU usage of FogMon is due to the fact that  it integrates resource-intensive monitoring probes (e.g., bandwidth testing), whereas we utilize the lightweight "psutil". A direct comparison of resource usage between DEMon and FogMon2 is not accurate since both tools employ different probes. DEMon collects limited metrics with a lightweight tool, and our focus is more on efficient communication overlay, storage, and retrieval. Existing probes could be embedded for more accurate measurements and a rich set of monitoring metrics.

\revisedmajor{}{
\textbf{Answer to RQ2:} 
DEMon exhibits a significantly lower overhead in resource consumption, making it highly suitable for resource-constrained edge environments. For example, it maintains an average CPU usage of less than 0.75\%. Additionally, the node bandwidth significantly decreases once the system converges. DEMon also has a smaller memory footprint, with the flexibility to adjust the runtime memory size according to specific requirements. }

\subsubsection{$RQ_3$: Query Latency}
To investigates the number of messages required to retrieve monitoring data about any particular node, we performed 100 queries each time for different failure rates. In this setting, each query requests the utilization metrics of a random node. The querying process follows the logic explained in the Algorithm \ref{alg:query-client}, where   \verb+quorum_number+ is set to 3, similar to popular distributed databases such as Cassandra.

We generate parallel queries to three random nodes for each query request. Once a node receives a request, it checks the local object store based on the requested node ID (i.e., IP address) as key and sends the monitoring data in the format described earlier (see Section \ref{subsec:state-repository}). We set different failure rates from 0\%-90\%  with the interval of 10\%. We randomly disconnect the \verb+failure_rate + \% of existing nodes for each failure rate setting. Despite this, DEMon can remarkably provide the information of a requested node without query failures. This is because, since every node stores information of every other node, even when 90\% of the nodes fail, the queried node information is successfully retrieved. In our results, we received the queried information with a maximum number of messages of 148 and a minimum number of messages of 3 (best case scenario, equal to \verb+quorum_number+), with an average value of 10.45 and a median of 3.

\revisedmajor{}{
\textbf{Answer to RQ3:}
DEMon offers a resilient framework capable of handling a high number of node failures. It ensures reliable data retrieval from decentralized nodes even in scenarios where up to 90\% of the nodes have failed.
}

\subsubsection{$RQ_4$: Age of Information (AoI)}

\begin{figure*}
  \captionsetup{justification=centering}
  \begin{subfigure}[t]{0.45\textwidth}
        \includegraphics[width=\linewidth]{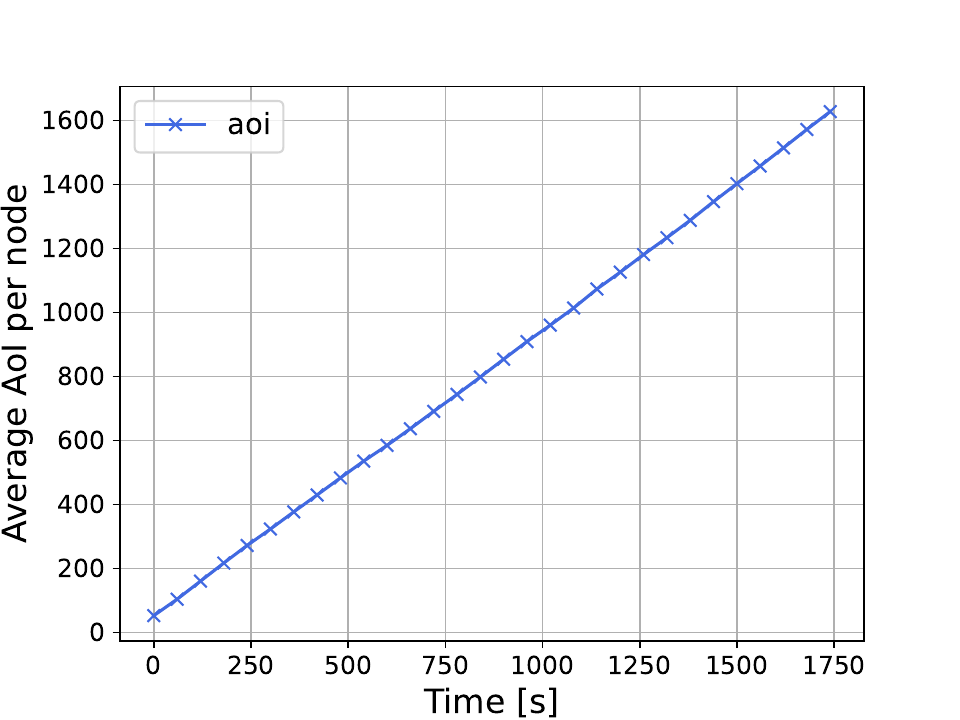} 
         \caption{Age of Information analysis (gossip\_count = 3).}
        \label{fig:aoi}
    \end{subfigure}
    \begin{subfigure}[t]{0.45\textwidth}
        \includegraphics[width=\linewidth]{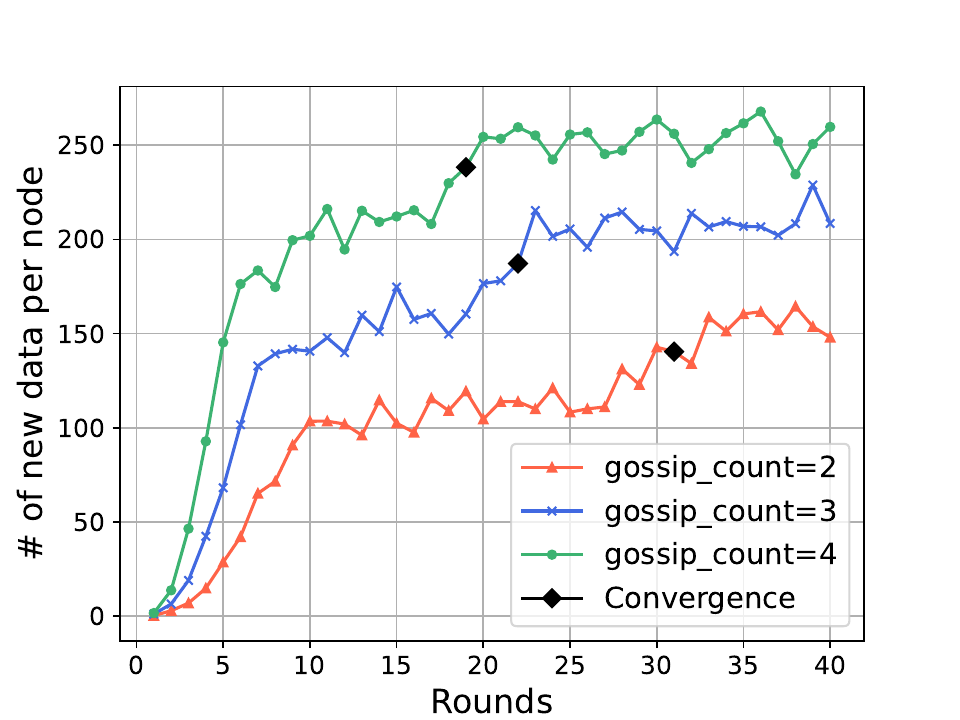} 
        \caption{New updated data of other nodes received (black diamond marker denotes the initial system convergence). }
        \label{fig:gn_rounds_freshdata}
    \end{subfigure}
    \caption{Analysis of AoI and new data updates (system size=300, gossip\_rate=1). }
    \label{fig:freshdata}
\end{figure*}

Although DEMon provides monitoring information of all nodes even when a high \% of nodes fail, it is essential to know the freshness of the data in all nodes, which is also a valuable aspect of monitoring data. We analyze the average \verb+AoI+ of all nodes for different time intervals. In this experiment, we take a snapshot of the monitoring data from all nodes every minute for a total of 30-minute intervals. Once the system is converged initially, the \verb+AoI+ of each node is calculated based on Equation \ref{eq:aoi}.

Figure \ref{fig:aoi} shows the overall \verb+AoI+ of the system (average from all the nodes) for all intervals. It can be observed that \verb+AoI+ of the overall monitoring system gradually increases as the monitoring time interval increases. This is expected since every node gossips to a very few nodes for every \verb+gossip_rate+ seconds; it would take several further gossip rounds to reach the updated information to other nodes, while each node simultaneously keeps on updating their monitoring information, creating a time delay between the two data instances. However, as Figure \ref{fig:gn_rounds_freshdata} shows, even after initial system convergence, all the monitoring agents receive other nodes' newly updated monitoring information in each round. 

The rate of new updates is highly dependent on the \verb+gossip_count+. When \verb+gossip_rate+ is set to 4, the average number of new fresh data updates is almost equal to the system size (i.e., 300).
Since DEMon is designed to hold multiple snapshots of a node's data in different nodes. If the most recent or fresh data is needed, higher \verb+gossip_count+ values can be configured.

\revisedmajor{}{
\textbf{Answer to RQ4:}
The results demonstrate that all edge nodes consistently receive new updates from their peers. Furthermore, DEMon offers controllable parameters that allow for the management of data freshness throughout the edge network.
}
\subsection{Use Case Evaluation on an RPi-based Testbed}

\begin{figure}[ht]
    \captionsetup{justification=centering}
    \centering
    \includegraphics[width=0.9\linewidth]{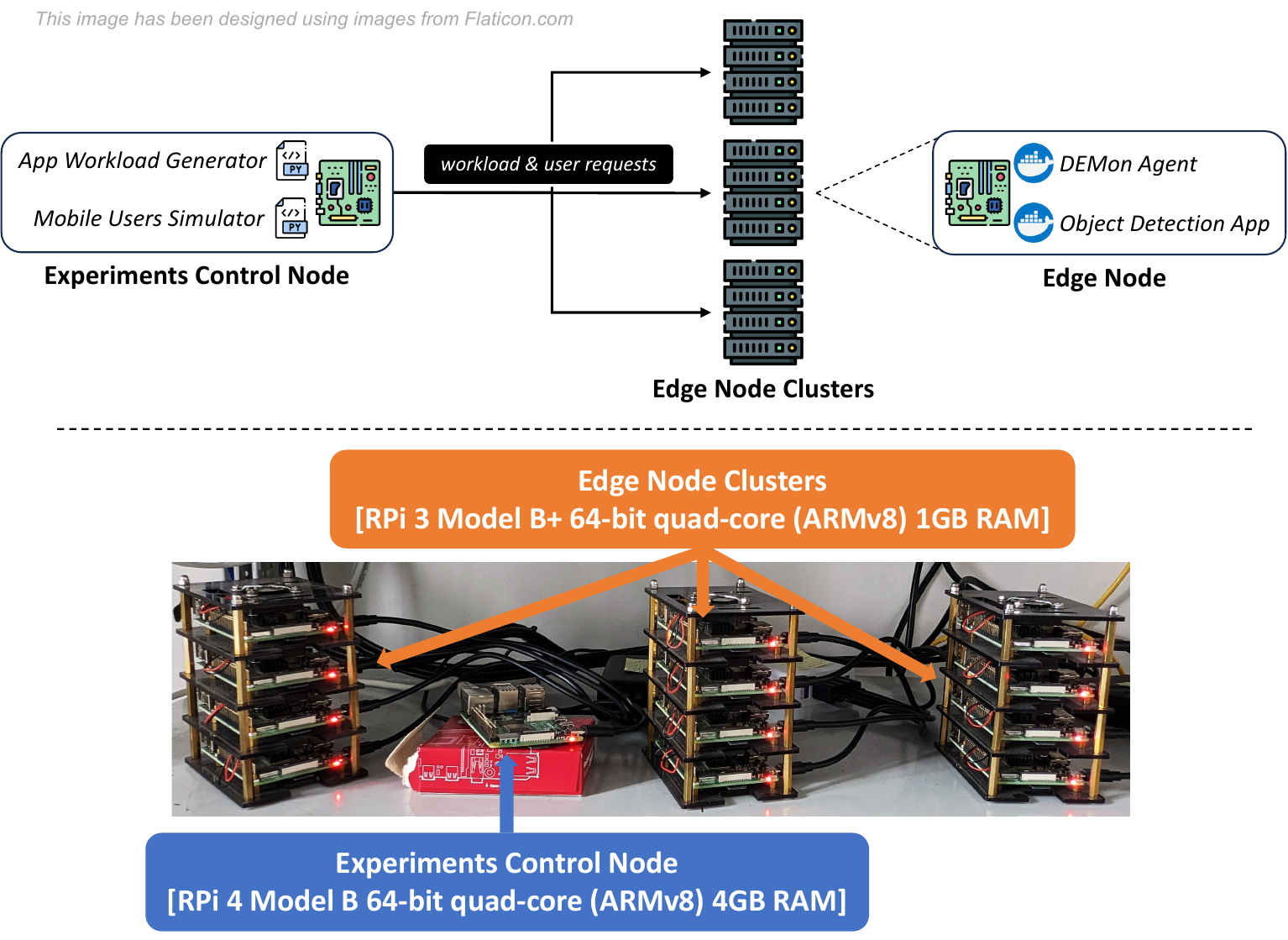} 
    \caption{RaspberryPi testbed for the use case evaluation with three Edge Clusters and an Experiments Controller node.}
    \vspace{-7 pt}
    \label{fig:testbed_rpi}
\end{figure}

In this section, we showcase the implementation of the use case described in Section \ref{sec:motivation}. Our primary objective is to provide fast and trustworthy monitoring information for near real-time mobile user applications deployed on top of volatile edge nodes.

Figure \ref{fig:testbed_rpi} depicts our Raspberry Pi based testbed, with 3 interconnected mini edge clusters (ECs), where each EC has 4 Raspberry Pi (RPi) nodes, for a total of 12 edge nodes. We also setup an additional RPi controller node to run our experiments. All the experimental material to reproduce the use-case is available in our publicly released repository\footnote{\repo}.

We emulate the geographical location of edge nodes and users based on the Edge User Allocation dataset \cite{lai2018optimal_mel_dataset}. This dataset provides the geo-locations of edge servers and users in the metropolitan area of Melbourne, Australia. We derive 12 edge server locations and associate their latitude and longitude values with our RPi nodes. Similarly, we generate the user mobility behaviour using the user location data from the same dataset, which has the location entries of users based on their mobile GPS location. We deploy the DEMOn Agent as a Docker container to all RPi nodes.

We use an object detection AI model as a representative workload in our IoT-Edge environment. We use a pre-trained YOLO object detection model  \footnote{https://pjreddie.com/darknet/yolo/} trained with the COCO dataset and deploy it on our Raspberry Pi cluster. The application takes a base64 encoded image as the input and returns a list of detected objects and bounding box dimensions of the objects as a response. We implement this application with a Python-based user application and deploy it as a Docker container. An instance of this application runs inside each edge node and exposes its services through HTTP APIs.

We generate the background workload using Locust, a workload generator tool \footnote{https://locust.io/}. We send multiple requests at an interval to all edge nodes with random images from the test dataset. Each node in a cluster receives incoming requests from $(1, 4)$ number of concurrent users, randomly chosen, and each user generates a new request within an interval of $(1, 15)$ seconds. This is important to evaluate our edge monitoring system in a dynamic workload environment.

Finally, the overall use case workflow is set up as follows. We extract 100 random user location entries from the user location dataset, and for each user location, the task is to offload computationally expensive tasks (e.g., object detection) to the nearest edge node. To do that, we need to know the suitable edge nodes that satisfy the user requirements, such as the availability and usage level of the nearest edge node, so that the user application can offload its computational task. 

To accomplish this, DEMOn provides real-time monitoring information to user requests. In our setup, we set the different levels of failures, i.e., $failure\_rate= \lbrace 50,60,70,80,90 \rbrace$  to generate a volatile edge environment. Similarly, we select \% CPU usage as the QoS parameter, as higher CPU usage would have higher computational latency for the offloading task and the lower the CPU threshold indicates the stricter QoS requirement (application task demands a node with low CPU usage). We set CPU QoS from $50-100\%$ with an interval of 10\%.

\begin{figure*}
\centering
  \captionsetup{justification=centering}
   \begin{subfigure}[t]{0.49\textwidth}
        \includegraphics[width=\linewidth]{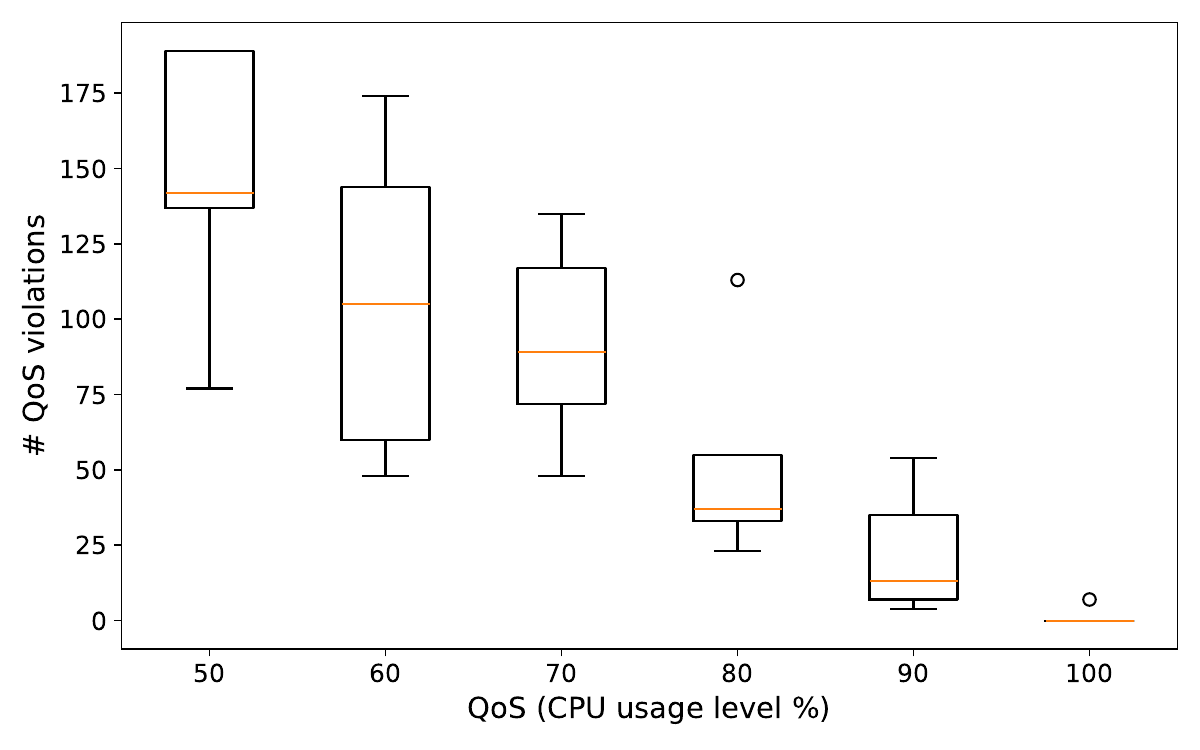} 
       \caption{CPU QoS vs QoS violations.}
        \label{fig:use_case_qos_violation}
    \end{subfigure}
    \begin{subfigure}[t]{0.49\textwidth}
        \includegraphics[width=\linewidth]{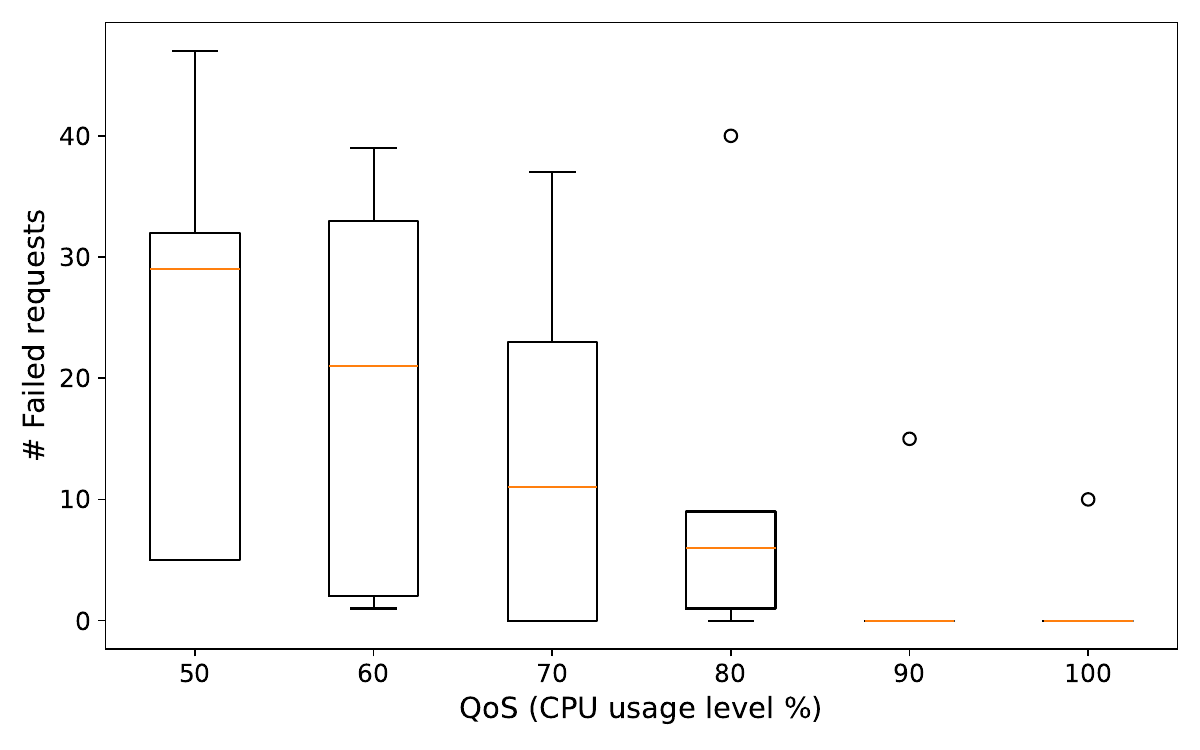}
        \caption{CPU QoS vs failed requests.}
    \label{fig:use_case_failed_requests}
   \end{subfigure}
       \caption{Object detection offloading use case: Application performance with different level QoS settings.}
        \label{fig:usecase_exp}
\end{figure*}

Figure \ref{fig:usecase_exp} shows the performance of offloading user requests to nearby edge nodes. We measure two important metrics: (1) \textit{Number of QoS violations} - indicating how many times QoS is potentially violated, i.e., the nearest node has higher CPU usage than the expected QoS, forcing us to offload to the next nearest node. (2) \textit{Number of failed requests} - indicating how many requests never find a suitable node in the edge network, thus, discarding the offloading requests. For both metrics, we present the average results across different failure rates configured as explained above. As seen in Figure \ref{fig:use_case_qos_violation}, when we have relaxed QoS (i.e., CPU threshold is 100\%), the offloading is mostly done to the nearest node regardless of its usage level, without requiring our request to be redirected to the next nearest node. However, when we have strict QoS, a higher number of QoS violations are found. This means offloading is done to the next nearest suitable node in the edge environment. For instance, with a CPU threshold of 50\%, the number of QoS violations had min=77, max=189, and mean=146.80. It is important to note that, to make such a decision, only one monitoring query is made to the nearest edge node initially. In fact, DEMon provides monitoring information of other nodes that are in the network. In other words, if no such monitoring system is present, we would have to make 147 extra queries to individual nodes in the worst case, while our system require one query.

Figure \ref{fig:use_case_failed_requests} shows the number of failed requests with different levels of QoS. The strict QoS setting (CPU threshold of 50\%) results in an extremely high number of failed requests, with min=5, max=47, and mean=23.60. Similarly, when we set relaxed QoS (CPU threshold of 100\%), the least number of failed requests are observed, with min=0, max=10, and mean=2.  In summary, the results obtained by the use case implementation show how DEMon enables efficient retrieval of decentralized monitoring data, supporting critical applications in volatile edge environments.

\subsection{\textcolor{black}{Threats to Validity}}
\revisedmajor{}{
In this work, we aim to establish a framework for deploying decentralized monitoring systems within highly volatile edge environments. We anticipate the following threats to our study. First, we have developed a prototype of our framework and conducted evaluations through emulation and on a small-scale testbed. However, this implementation is not directly transferable to real-world deployments due to additional critical components that must be considered. These include establishing secure communication  (such as authentication and authorization) and creating static bootstrap nodes within the network, among others. Second, for the sake of rapid prototyping and ease of use,  we chose Python and REST APIs for implementation and method evaluation. We recognize that further performance enhancements could be achieved by opting for more resource-efficient programming languages (e.g., C/C++ or Go) and communication frameworks (e.g., gRPC). 
Finally, to mitigate statistical bias in our results, we repeated each experiment three times and reported the mean value for each data point. Additionally, we have included standard deviations in the plots, where applicable to provide a measure of variability.

Nevertheless, our methodological approach proves that DEMon is particularly effective in highly volatile edge environments where failures are common and reliable centralized monitoring is not feasible.
}

\section{Related Work}
\label{sec:related-work}

\begin{table}[h]

\caption{
{\revisedmajor{}{A comparison of related works}
}
}
\label{table:relatedwork}
   \fontsize{20pt}{20pt}\selectfont
\setredmajor
\resizebox{\columnwidth}{!}{
\setredmajor
\begin{tabular}{lllllll}
\hline
                       & \multicolumn{1}{c}{\textbf{System}} & \multicolumn{1}{c}{\textbf{Characteristics}}                                                                      & \textbf{Implementation/Evaluation Method}                                                                                            & \textbf{Communication Protocol}                                                                     & \multicolumn{1}{c}{\textbf{Storage Architecture}} & \multicolumn{1}{c}{\textbf{Trustworthiness}} \\ \hline
Wuhib et al. ~\cite{Wuhib_2007}           & G-GAP                               & Monitoring of network-wide aggregates                                                                             & \begin{tabular}[c]{@{}l@{}}Simulation using SIMPSON\\ simulator and traces\end{tabular}                                              & Gossip                                                                                              &      Centralized                                             & \xmark                        \\
Ward et al.   ~\cite{ward2013monitoring}         & N/A                                 & Monitoring IaaS cloud resources                                                                                   & Simulation                                                                                                                           & Layered Gossip                                                                                      & Partially centralized                             & \xmark                        \\
Taherizadeh et al. ~\cite{Taherizadeh_2017}    & N/A                                 & Network monitoring for real-time edge services                                                                    & Implemented with JCatascopia framework                                                                                               & N:1 communication                                                                                   & Centralized                                       & \xmark                        \\
Battula et al.   \cite{battula_scb_tsc20}       & SCB                                 & \begin{tabular}[c]{@{}l@{}}Fog monitoring using support confidence-based \\ technique\end{tabular}                & \begin{tabular}[c]{@{}l@{}}Implemented emulation tool in Java and use\\ case evaluation\end{tabular}                                 & \begin{tabular}[c]{@{}l@{}}Hierarchical regions (clusters) \\ with centralized server\end{tabular}  & Centralized                                       & \xmark                        \\
Grogman et al. ~\cite{ grogman_2017}         & PyMon                               & \begin{tabular}[c]{@{}l@{}}A tool to collect container statistics running \\ on heterogeneous nodes\end{tabular}  & Implementation using extending "monit" tool                                                                                          & N:1 communication                                                                                   & Centralized                                       & \xmark                        \\
Brandon et al.   ~\cite{brandon2018fmone}      & FMone                               & \begin{tabular}[c]{@{}l@{}}Monitoring of edge resources based on\\ use case-specific workflow\end{tabular}        & \begin{tabular}[c]{@{}l@{}}Prototype implementation and evaluation\\ on Grid5K environment\end{tabular}                              & \begin{tabular}[c]{@{}l@{}}Hierarchical regions (clusters) \\ with centralized servers\end{tabular} & Centralized                                       & \xmark                        \\
Colombo  et al. ~\cite{adaptievemon}           & AdaptiveMon                         & Self-adaptive P2P monitoring of Fog                                                                               & Implemented in C++ extending FogMon2                                                                                                 & \begin{tabular}[c]{@{}l@{}}P2P communication \\ with leader and followers\end{tabular}              & Partially centralized                             & \xmark                        \\
Forti et al.   ~\cite{fogmon, gaglianese2023assessing_fogmon2}        & FogMon                              & Distributed Monitoring for Fog infrastructures                                                                    & TRL5 implementation using C++                                                                                                        & \begin{tabular}[c]{@{}l@{}}P2P topology \\ with leaders and followers\end{tabular}                  & Partially centralized                             & \xmark                        \\
\textbf{Our Work} & \textbf{DEMon}                      & \textbf{\begin{tabular}[c]{@{}l@{}}A completely decentralized monitoring\\ system for volatile edge\end{tabular}} & \textbf{\begin{tabular}[c]{@{}l@{}}Prototype implementation with Python, \\ evaluation in emulation and use case setup\end{tabular}} & \textbf{Gossip}                                                                                     & \textbf{Decentralized}                            & \textbf{\cmark}               \\ \hline
\end{tabular}
}

\end{table}

Traditional monitoring systems in distributed computing are mainly centralized systems, where a set of dedicated servers periodically collect data from all the resources either through \textit{push}  or \textit{pull} based APIs. Cloud monitoring tools such as Google's Borgmon~\cite{beyer2016site}, or Prometheus~\cite{turnbull2018monitoring} provide flexibility to store data on local disk or remote storage; their control plane (e.g., alert management and query processing) is still centralized. Similar design principles are followed in  High-Performance Computing (HPC) monitoring systems. However, existing Edge Computing platforms depend on a centralized monitoring system consequent to the design principle of edge systems. For instance, Kubernetes \footnote{https://kubernetes.io} and its variants, such as KubeEdge, which are popularly used in edge infrastructures for resource and application management, use a centralized control plane and monitoring system. Such approaches are infeasible in critical edge infrastructures requiring strict latency and reliability under failures.


\par \textbf{Edge  monitoring.} 
Researchers have made many efforts to address the challenges in Edge Computing monitoring. Grogman et al.~\cite{ grogman_2017} propose PyMon, a monitoring system targeting container-based computing architectures with a small resource footprint. The solution is primarily targeted at IoT-based single-board edge devices. The system was built as a set of container images providing monitoring service.
Similarly, Taherizadeh et al~\cite{Taherizadeh_2017} propose a network monitoring approach for data streaming applications, where each edge node hosts a monitoring probe and pushes the data to a centralized time-series database servers. It considers QoS metrics such as delay, packet loss, throughput, and jitters important for streaming applications.
Brandon et al.~\cite{brandon2018fmone} propose FMonE, designing a monitoring solution that considers elasticity and resiliency, addressing the unique challenges of edge systems. The prototype system uses Docker containers to build monitoring agents. However, storage and information processing depend on centralized database systems for specific application domains. All these solutions, including FMonE, either aggregate data from several edge sites with a central controller, or push monitored data to the Cloud. 

Other works explored self-adaptive monitoring for multi-tier Fog computing systems. In particular, Forti et al.~\cite{fogmon} propose FogMon, a hierarchical P2P monitoring system where lower-tier nodes are dedicated as Followers and higher-tier nodes as Leader nodes. The inherited self-adaptive P2P architecture characteristics make it suitable for volatile Fog-Edge environments.
Similarly, Colombo and Tundo, et al.~\cite{adaptievemon} propose AdaptiveMon, a self-adaptive P2P monitoring system that can dynamically change its behavior according
to collected monitoring indicators by leveraging a rule-based expert system. Nevertheless, both these approaches still exhibit a partially centralized architecture in the form of Leader nodes, and they are only feasible when multi-tier deployments exist with Edge, Fog, and Cloud nodes.
\revisedmajor{}{
Battula et al. \cite{battula_scb_tsc20} proposed a fog monitoring framework that utilizes a Support and Confidence-Based (SCB) technique to optimize resource usage of monitoring services. Their method takes into account the contextual information of resources, such as the current battery power of fog devices, to determine their participation in the monitoring service. The use case is assessed by grouping fog devices under a fog leader, with leaders sharing the monitoring data with a centralized server.

Gaglianese et al. \cite{gaglianese2023assessing_fogmon2, fed4fire_22} extended the FogMon \cite{fogmon} framework, adding new features and providing a TRL5 level prototype system for monitoring fog devices. This extension follows FogMon's architecture for information dissemination and offers improved accuracy in monitoring data and fault resiliency with improved management of different types of failures. The proposed approach was evaluated on up to 40 edge nodes deployed on the Fed4Fire testbed. An extensive list of monitoring services in the fog environment and their comparative study is presented in \cite{monitoring_survey_cn_22}. These studies rely on partially centralized architectures that are suitable for the fog environment with reasonable reliability. We provide a brief comparison of the most relevant related work in Table \ref{table:relatedwork}. As observed, our work adopts a completely decentralized approach in the design of the monitoring service for the edge environment, ensuring trustworthy retrieval of monitoring data from distributed storage.
}

\textbf{Gossip protocol in monitoring.} 
Gossip protocols have wide applications in many domains. Especially various studies have used the Gossip protocol for monitoring networks and distributed systems. Ward et al.~\cite{ward2013monitoring} propose a solution to monitor large-scale cloud systems with layered Gossip protocols. The authors group the resources into multiple layers, from a group of virtual machines to multi-region data centers. The multi-layered inter-group and inter-cloud communication protocol architecture are considered to match the bandwidth requirement, i.e., high-speed network connection within a data center, less reliable, and low-speed networks available between data center regions. The Gossip protocol is used to spread the information across all the virtual machines. However, how the information is stored and retrieved in the system is unclear. Similarly, Van et al.~\cite{van2003astrolabe} propose Astrolabe, one of the earlier systems for distributed information management based on the Gossip protocol. It is mainly designed for resource monitoring, management, and data mining. With the combination of P2P Gossip protocol, mobile code, and SQL query language, the system is implemented to manage the data collection, storage, and aggregation in real-time. It organizes resources in hierarchical domains (e.g., Domain Name System. 
Furthermore,  Wuhib et al.~\cite{Wuhib_2007} use the Gossip protocol to monitor network-wide aggregate metrics (e.g., average, min, and max). Gossip-based protocols are used in other distributed systems tasks such as in failure detection \cite{van1998gossip} and network size and density estimation \cite{DARWISH2015337}. In these works, the main focus is on aggregating the information of resources and they do not address a wider spectrum of monitoring requirements, including information dissemination and retrieval methods. \revisedmajor{}{The decentralized databases such as Cassandra use the gossip protocol for implementing internode communication within data centers. However, these distributed databases still operate with a centralized control plane for communication coordination. Similarly, other decentralized databases, including weavedb\footnote{https://github.com/weavedb/weavedb} and orbited \footnote{ https://github.com/orbitdb/orbitdb}, are heavyweight since they use compute-heavy logics like blockchains and are mainly designed to store application business data. Thus, these Off-the-shelf databases are unsuitable for edge monitoring systems.
}

\par However, the primary objective of our work is to provide a framework and techniques for a self-adaptive, self-configurable, and trustworthy monitoring system applicable to volatile edge environments, without limiting to specific application tasks. The individual components, such as querying languages of related work, act as complementary solutions in our approach. 

\section{Conclusions and Future Directions}
\label{sec:conclusions}

Self-adaptive and decentralized monitoring systems are necessary for efficient management of the edge environments. However, existing approaches are centralized in architecture, which increases information storage and retrieval latency and creates failure bottlenecks, making them infeasible for highly-volatile edge environments.

In this work, we presented DEMon, a monitoring system designed to operate autonomously without any external controller and configuration manager, which stores monitoring data in a decentralized fashion. DEMon enables efficient information spreading by leveraging a Gossip-based protocol, and it provides a trustable retrieval mechanism for distributed information with a leaderless quorum consensus technique. Additionally, it allows hyper-parameter configuration based on the system properties. We implemented the proposed framework as a lightweight and interoperable container-based system, and we performed an extensive evaluation with both a simulated large-scale edge environment and an in-lab testbed. Our experimental results show that DEMon quickly spreads the monitored information, and can retrieve monitoring information even when most of the nodes fail. 

In the future, we plan to extend the query capabilities of the system by integrating a query language for extracting granular data.


\bibliographystyle{ACM-Reference-Format}
\bibliography{bibiliography}

\end{document}